\documentclass[useAMS,usenatbib,usedcolumn]{mn2e}
\def\HI{\ifmmode{\rm HI}\else{H\/{\sc i}}\fi}

\def\sun{\hbox{\scriptsize $\odot$}}
\def\lsun{\ifmmode{{\mathrm L}_{\odot}}\else{L$_{\odot}$}\fi}

\def\msun{\ifmmode{{\mathrm M}_{\odot}}\else{M$_{\odot}$}\fi} 
\def\msunpc2{\ifmmode{{\mathrm M}_{\odot} \, {\mathrm{pc}}^{-2}}\else{M$_{\odot} \, {\mathrm {pc}}^{-2}$}\fi}

\def\kms{\ifmmode{{\mathrm{km \, s^{-1}}}}\else{${\mathrm{km \, s^{-1}}}$}\fi}

\def\la{\mathrel{\mathchoice {\vcenter{\offinterlineskip\halign{\hfil
$\displaystyle##$\hfil\cr<\cr\sim\cr}}}
{\vcenter{\offinterlineskip\halign{\hfil$\textstyle##$\hfil\cr
<\cr\sim\cr}}}
{\vcenter{\offinterlineskip\halign{\hfil$\scriptstyle##$\hfil\cr
<\cr\sim\cr}}}
{\vcenter{\offinterlineskip\halign{\hfil$\scriptscriptstyle##$\hfil\cr
<\cr\sim\cr}}}}}
\def\ga{\mathrel{\mathchoice {\vcenter{\offinterlineskip\halign{\hfil
$\displaystyle##$\hfil\cr>\cr\sim\cr}}}
{\vcenter{\offinterlineskip\halign{\hfil$\textstyle##$\hfil\cr
>\cr\sim\cr}}}
{\vcenter{\offinterlineskip\halign{\hfil$\scriptstyle##$\hfil\cr
>\cr\sim\cr}}}
{\vcenter{\offinterlineskip\halign{\hfil$\scriptscriptstyle##$\hfil\cr
>\cr\sim\cr}}}}}

\def\aj{AJ}%% Astronomical Journal
\def\araa{ARA\&A}%% Annual Review of Astron and Astrophys
\def\apj{ApJ}%% Astrophysical Journal
\def\apjl{ApJ}%% Astrophysical Journal, Letters
\def\apjs{ApJS}%% Astrophysical Journal, Supplement
\def\aap{A\&A}%% Astronomy and Astrophysics
\def\mnras{MNRAS}%% Monthly Notices of the RAS
\def\pasa{PASA}%% Publications of the Astron. Soc. of Australia
\usepackage[figuresright]{rotating}
\usepackage{lscape}
\usepackage{graphics}
\usepackage{epsfig}
\usepackage{multirow}
\usepackage{bigdelim}
\usepackage{bigstrut}
\usepackage{amsmath}
\usepackage{amssymb}
\usepackage{lscape}
\usepackage{supertabular}
\usepackage[3D]{movie15}
\usepackage{hyperref}
\usepackage{epstopdf}

\defcitealias{Jaffe2012b}{J12b}

\title[BUDHIES I]{BUDHIES I: characterizing the environments in and around two clusters at z$\backsimeq$0.2 }
\author[Y.~Jaff\'e et al.] {Yara L. Jaff\'e$^{1,2,3}$\thanks{E-mail: yara.jaffe@astro-udec.cl}, Bianca M. Poggianti$^{2}$, Marc A. W. Verheijen$^{3}$, Boris Z. Deshev$^{3,4}$, \and
Jacqueline H. van Gorkom$^{5}$\\ 
   $^1$Department of Astronomy, Universidad de Concepci\'on, Casilla 160-C, Concepci\'on, Chile\\
   $^2$INAF - Osservatorio Astronomico di Padova, vicolo dell' Osservatorio 5, I-35122 Padova, Italy\\
   $^3$University of Groningen, Kapteyn Astronomical Institute, Landleven 12, 9747 AD, Groningen, The Netherlands\\
   $^4$Tartu Observatory, T\~oravere, 61602, Estonia\\
   $^5$Department of Astronomy, Columbia University, Mail Code 5246, 550 W 120th Street, New York, NY 10027, USA}
\begin{document}

%%\date{v1.2; 26-3-2008}

\maketitle

\begin{abstract}

We present the optical spectroscopy for the Blind Ultra Deep HI Environmental Survey (BUDHIES). 
With the Westerbork Synthesis Radio Telescope, BUDHIES has detected HI in over 150 galaxies in and around two 
Abell clusters at $z\simeq0.2$. 
With the aim of characterizing the environments of the HI-detected galaxies, 
we obtained multi-fiber spectroscopy with the William Herschel Telescope.
In this paper, we describe the spectroscopic observations, report redshifts and EW[OII] measurements for $\sim600$ galaxies, and perform an environmental analysis. In particular, we present cluster velocity dispersion measurements for 5 clusters and groups in the BUDHIES volume, as well as a detailed substructure analysis. 
%The spectroscopic data presented in this paper will be used in future papers to explore the effect of environment (as well as stellar mass) 
%%The results shown here will be used in a series of forthcoming papers to explore the effect of environment and stellar mass 
%on the HI-gas, morphologies and star-formation in BUDHIES. %galaxies at $z\simeq0.2$. 

\end{abstract}


%%%%%%%%%%%%%%%%%%%%%%%%%%%%%%%%%%%%%%%%%%%%%%%%%%%%%%%%%%%%%%%%%%%%%%%%%%%%%%%
%                                                                             %
%  1. Introduction                                                            %
%  \label{sec:introduction}                                                   %
%                                                                             %
%%%%%%%%%%%%%%%%%%%%%%%%%%%%%%%%%%%%%%%%%%%%%%%%%%%%%%%%%%%%%%%%%%%%%%%%%%%%%%%
\section{Introduction}
\label{sec:introduction}

Rich clusters of galaxies, in relation to the large scale structure in which they are embedded, offer
a unique laboratory to study the effects of global and local environments on the properties of their
constituent galaxies. Evidence has accumulated that the star formation activity, as well as galaxy morphology 
strongly depend on the environment in which galaxies are located. 
This is well exemplified by the morphology-density relation \citep{Dressler1980}, 
that shows that early-type galaxies are more frequent in regions with high local density 
(the number of galaxies per unit projected area, or volume), while spirals dominate the low-density regions.

The evolution of galaxy properties in clusters is strong, even in the last few Gyrs: the 
fraction of blue galaxies in clusters was higher in the past \citep[Butcher-Oemler B-O effect,][]{bo78}, and the relative number of S0  galaxies increases with time, at the expense of the spiral population \citep{Dressler1997,vanDokkum1998,fasano2000,Desai2007}. 
Similar trends are now known to take place also in the field \citep[]{Bell2007, Oesch2010}, where the relative numbers of  red, passively evolving, early-type galaxies increases with time.

Recent large surveys \citep[SDSS, 2dF,][]{lewis02,gomez03,Haines2007,Mahajan2012} and studies of groups, cluster outskirts and filaments \citep[e.g.][]{Treu2003, Wilman2009, Fadda2008, Porter2008, Roychowdhury2012} have shown that the environmental dependencies extend to the lowest density environments. 
It is now thought that galaxies may be ``preprocessed'' before they fall into clusters, i.e. environmentally driven evolution occurs in lower density environments. 
A remaining question is where and how does this happen. 
% However, many issues remain unsolved. For instance, to what extent the evolution observed in clusters is driven by processes occurring in the cluster outskirts or in the surrounding filaments and groups?, or to what extent
%galaxy evolution on a cosmic scale, in the general field, is driven by environmental processes? 

\begin{figure*}
\begin{center}
  \includegraphics[width=0.80\textwidth]{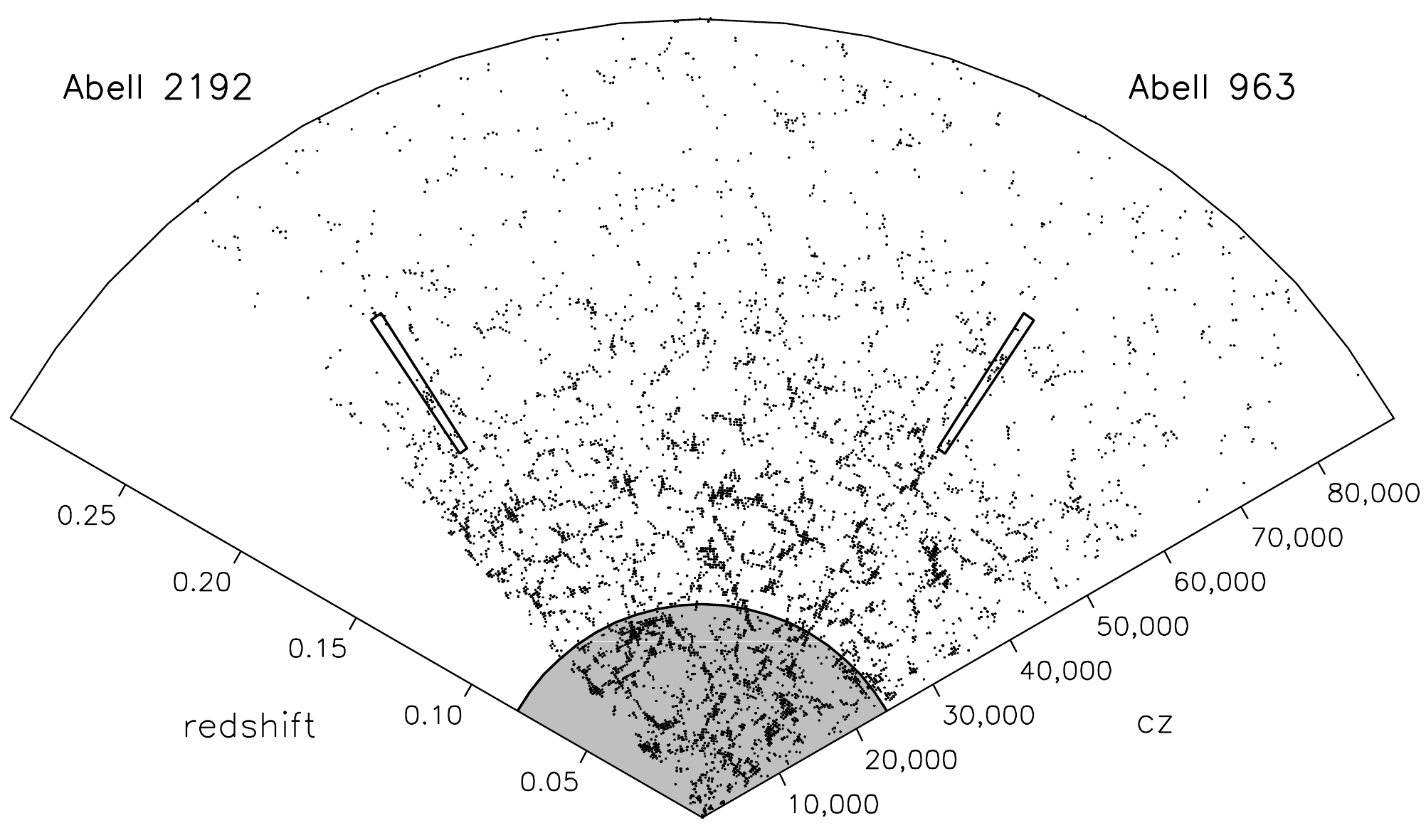}
\end{center} 
\caption{A redshift pie-diagram. Dots represent galaxies with optical redshifts from the SDSS. The elongated rectangles indicate the two volumes that have been surveyed with the WSRT (at $0.164 \leqslant z \leqslant 0.224$) . At these redshifts, the large scale structure becomes sparsely sampled by the SDSS. 
 The lower light-gray ``pie'' indicates the highest redshift ($z=0.08$) out to which HI imaging data existed before. 
Note that the HIPASS survey \citep{Meyer2004} does not extend beyond cz=12,000 km s$^{-1}$ while the Alfalfa survey \citep{Giovanelli2005} with
the Arecibo telescope is restricted to $z\lesssim0.06$.}
 \label{zpie}
\end{figure*}

Further clues (or perhaps questions) come from  the structure formation scenario of $\Lambda$ cold dark matter ($\Lambda$CDM) cosmology, that predicts that many galaxies have undergone the transition from field to cluster environments since z $\lesssim$ 1 \citep{delucia2012}. In fact, there is an ongoing debate on whether it is the cluster environment that drives galaxy evolution and transforms galaxies (nurture),  the field population that accretes onto clusters and that evolves with cosmic time (nature), or both \citep[e.g.][]{Poggianti1999,Kauffmann1999,KodamaBower2001,Ellingson2001,Desai2007,Bolzonella2010,Vulcani2010,Jaffe2011b}. 

A crucial tracer of galaxy evolution is the neutral atomic hydrogen gas from which the stars are formed. HI is a sensitive tracer of different environmental processes, in particular of tidal interactions and ram-pressure stripping. 
Observational evidence \citep[e.g.][]{Cayatte1990,BravoAlfaro2000,BravoAlfaro2001,PoggiantiVG2001,Kenney2004,crowl05a,Chung2007,Chung2009,Abramson2011,Scott2010,Scott2012} have suggested that the HI gas is disturbed and eventually truncated and exhausted in the cores of galaxy clusters, where the intra-cluster medium is denser. 
This is further supported by simulations \citep[see][for a review]{Roediger2009}, that have suggested that the cause of the distortion is ram pressure stripping and gravitational interaction with the cluster \citep[e.g. ][and references therein]{Vollmer2003,TonnesenBryan2009,Kapferer2009}. 
Moreover, observational studies have found that HI distributions showing interaction and gas stripping are particularly common in galaxy groups \citep[e.g.][and references therein]{VerdesMontenegro2001,vanDerHulst1979,Kern2008,Freeland2009,Hibbard2001}. 
In particular, \citet{VerdesMontenegro2001} showed that the HI content of compact groups is a decreasing function of their compactness.

Due to technological limitations, studies of the neutral atomic hydrogen gas content in galaxies have been mostly carried out in the local Universe ($z<0.08$).
The main impediments for HI surveys to be carried at higher redshifts  are the necessarily long integration times and the occurrence of man-made interference at those frequencies.

With the aim of understanding where, how, and why star-forming spiral galaxies get transformed 
into passive early-type galaxies, we have embarked on a Blind Ultra Deep HI Environmental Survey (BUDHIES) with the  
Westerbork Synthesis Radio telescope (WSRT).  
We refer to \citet[][]{Verheijen2007} and Deshev et al. (in preparation) for technical details on the HI survey. The strategy has been to study in detail two galaxy clusters, at z$\simeq$0.2, and the large scale structure around them.  
The surveyed clusters Abell 2192 and Abell 963 (A2192 and A963 from now on) are very different. A963 is a massive lensing cluster with an unusually large fraction of blue galaxies \citep{Butcher1983}. A2192 is a less massive cluster with significant substructure \citep{Jaffe2012}. 
Our aim is to identify the physical  mechanisms (e.g. tidal interactions, gas stripping, etc) that govern these transformations. 
The unique aspect of our study is that we have accurate measurements of the HI gas content.  While much optical work has been done on clusters at intermediate redshifts (z=0.4 and above),  and much HI work on nearby clusters (z$<$0.08), our study is the first where optical properties and gas content are combined at a redshift where evolutionary effects begin to show.  

We have surveyed the two clusters with several instruments in different wavelengths. At the core of the survey are the  ultra-deep HI observations, carried out at the WSRT, that yielded over 150 detections of galaxies with optical counterparts at z$\simeq$0.2. 
To assess the environments of the HI-detections, we have pursued optical spectroscopic observations, which are the subject of this paper. 

Our final goal is to relate the HI characteristics of the galaxies to their environment \citep[see first results in][]{Jaffe2012}, as well as to the
galaxy luminosities, stellar masses, morphologies, star formation rates and histories. In this paper, we  describe the spectroscopic observations carried out at the William Herschel Telescope (WHT), and the characterization of environment. 
We leave the rest of the analysis for a series of upcoming papers. 

In Section~\ref{sec:data} we summarize the survey strategy and the multi-wavelength data collected to date. In Section~\ref{sec:spec} we present the spectroscopic observations and the data reduction, the redshift and EW[OII] measurements, and the spectroscopic completeness.  In Section~\ref{sec:mem} we identify the clusters and groups in the sample and compute velocity dispersion, $R_{200}$ and cluster membership. We then inspect the clusters thoroughly for substructure in Section~\ref{sec:substructure}, and summarize the main characteristics of each of the identified cluster/groups in Section~\ref{sec:env_sum}. Finally, we present our conclusions in Section~\ref{sec:concl}.

Throughout this paper, we use Vega magnitudes and assume a ``concordance'' $\Lambda$CDM cosmology with $\Omega_{\rm M}=$0.3, $\Omega_{\Lambda}=$0.7, and H$_{0}=$70 km s$^{-1}$ Mpc$^{-1}$, unless otherwise stated.

\section{The BUDHIES project}
\label{sec:data}

The WSRT deep environmental survey targeted two clusters at $z\backsimeq0.2$, as well as the large scale structure around them.
The surveyed clusters, A2192 and A963, are very distinct. 
A963 at z=0.206 is a massive lensing B-O cluster with an unusually large fraction of blue galaxies \citep[$f_{\rm B}$=0.19, ][]{Butcher1983}, and a total X-ray luminosity of $L_{X}\simeq 3.4\pm1\times10^{44}h^{-2}$ergs/s \citep{Allen2003}.  %, and its regular X-ray contours are centered on a central cD galaxy, which suggests a low level of substructure.  
A2192 at z=0.188 is a less massive cluster in the process of forming, with a high degree of substructure \citep{Jaffe2012}. It has a velocity dispersion of 653 km s$^{-1}$ and is barely detected in X-rays \citep[$L_{X}\simeq7\times10^{43}h^{-2}$ erg/s; ][]{Voges99}. With the assumed cosmology, the spatial scales at the distances of A963 and A2192 are $3.4$ and $3.1$ kpc arcsec$^{-1}$ respectively.

After integrating for 117$\times12^{\rm h}$ on A963 and 76$\times12^{\rm h}$ on A2192 with the WSRT, more than 150 galaxies were detected and imaged in HI. The observations span a redshift range of $0.164-0.224$, a luminosity distance of $0.79-1.11$ Gpc, and a lookback time interval of $2.04-2.68$ Gyr. 
The results of a pilot study are presented in \citet[][]{Verheijen2007}. %Details on the WSRT observations can be found in Deshev et al. (in preparation). 
In short, this recently completed 4-year long-term Large Programme 
focused on two (single-pointing) volumes, each containing an Abell cluster of galaxies, as well as fore- and/or background voids, sampling the broadest range of cosmic environments. 
The depth of the WSRT observations allow a solid detection of a minimum HI mass of $2 \times 10^{9}$ M$_{\sun}$ (5 sigma), assuming a typical width of 100 km s$^{-1}$. 
It should be noted that the galaxy clusters only occupy $\sim$4\% of the surveyed 
volumes while the entire combined volume that has been blindly surveyed with the WSRT is 
$\sim 73,000$ Mpc$^{3}$, equivalent to the volume of the Local Universe within a distance of 26 Mpc.

In addition to the HI data, we have obtained near and far UV imaging with GALEX (Montero-Casta\~no et al. in preparation); 3.6, 4.5, 5.6, 8, 24 and 70 micron imaging with Spitzer (Cybulski et al. in preparation); Herschel  imaging (with SPIRE and PACS, Yun et al. in preparation); %CO observations with the NRO and LMT (Chung et al. in preparation); 
as well as 36 spectra with the Wisconsin Indiana Yale NOAO (WIYN) Telescope in the field of A2192. Finally, we also obtained $B$- and $R$-band imaging with the Isaac Newton Telescope (INT) in both clusters. These observations over a 1 square degree field were carried out under significantly better seeing conditions, and are much deeper than the available Sloan Digital Sky Survey (SDSS) photometry.

Figure~\ref{zpie} shows a redshift pie-diagram from available SDSS redshifts, showing the location of our surveyed clusters at $z\simeq0.2$ and highlighting the depth of the HI survey. This diagram already shows overdensities at the studied redshifts but it is evident that additional spectroscopy is needed to define the structures. 
In addition to the SDSS and WIYN redshifts, we also have 89 redshifts in the very centre of A963 from \citet[][and private communication]{LaveryHenry1994},  as well as 111 redshifts from Oliver Czoske (private communication) in the same field. However, the redshifts from \citet{LaveryHenry1994} were used cautiously, as they suffer from large uncertainties.

In the following section we present optical spectroscopy of a large number of galaxies in the two volumes, obtained at the William Herschel Telescope (WHT). 

%%%%%%%%%%%%%%%%%%%%%%%%%%%%%%%%%%%%%

\section{Optical Spectroscopy}
\label{sec:spec}

\subsection{Target selection}
\label{subsec:target}

To design the spectroscopic observations, we made use of the $B-$ and $R-$band magnitudes, the astrometry, and the galaxy-star separation from the INT photometry, as well as the HI-detection information and the available redshifts.  
Our selection and prioritization criteria are summarized in the following:

\textbf{(i) Type of object:}  the galaxy-star separation index (as given by \textit{SExtractor}) in either the $B$- or the $R$-band was $\leq 0.3$, in order to clean the sample from objects that are not galaxy-like. 

\begin{figure*}
\begin{center}
  \includegraphics[width=0.49\textwidth]{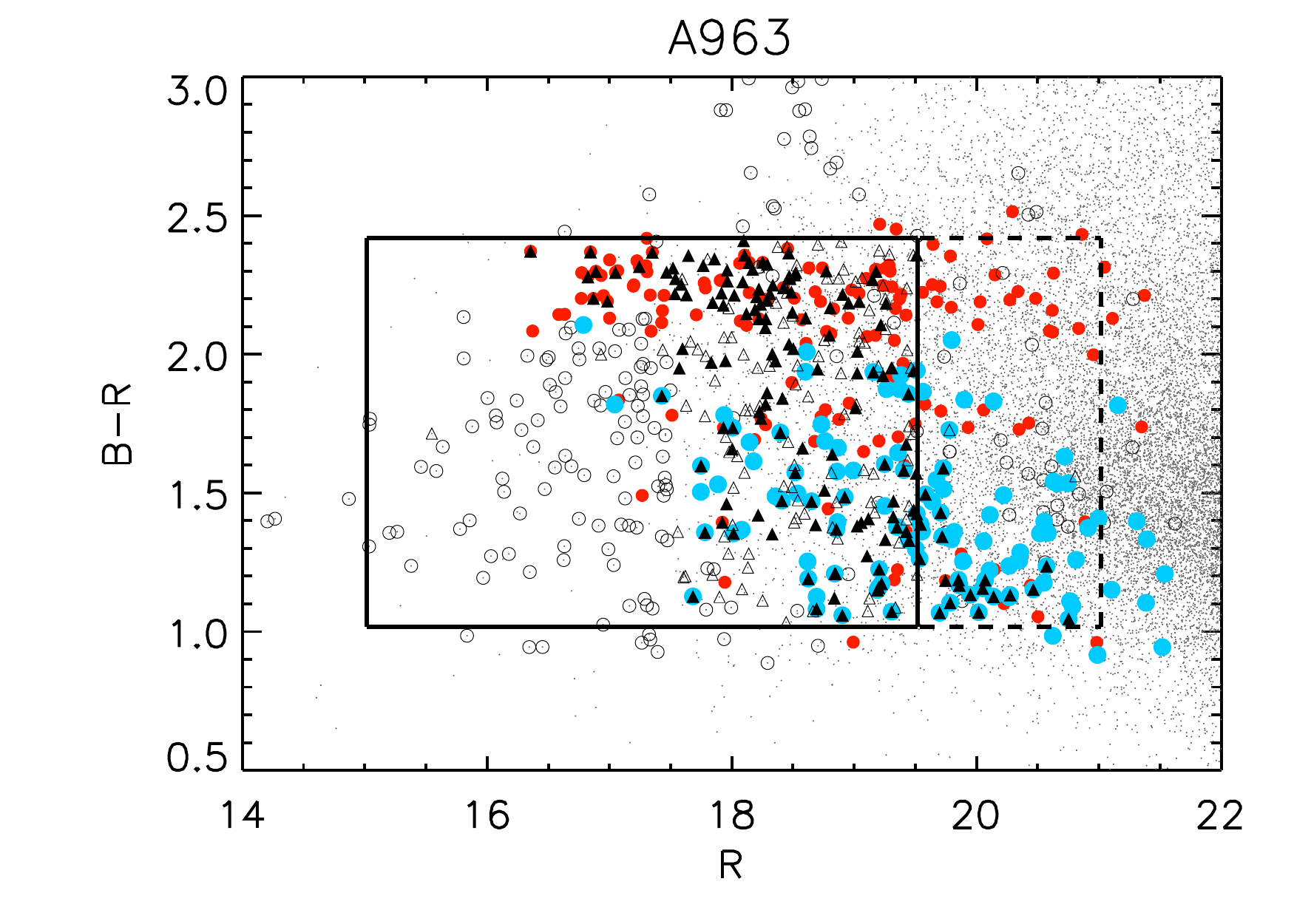}
  \includegraphics[width=0.49\textwidth]{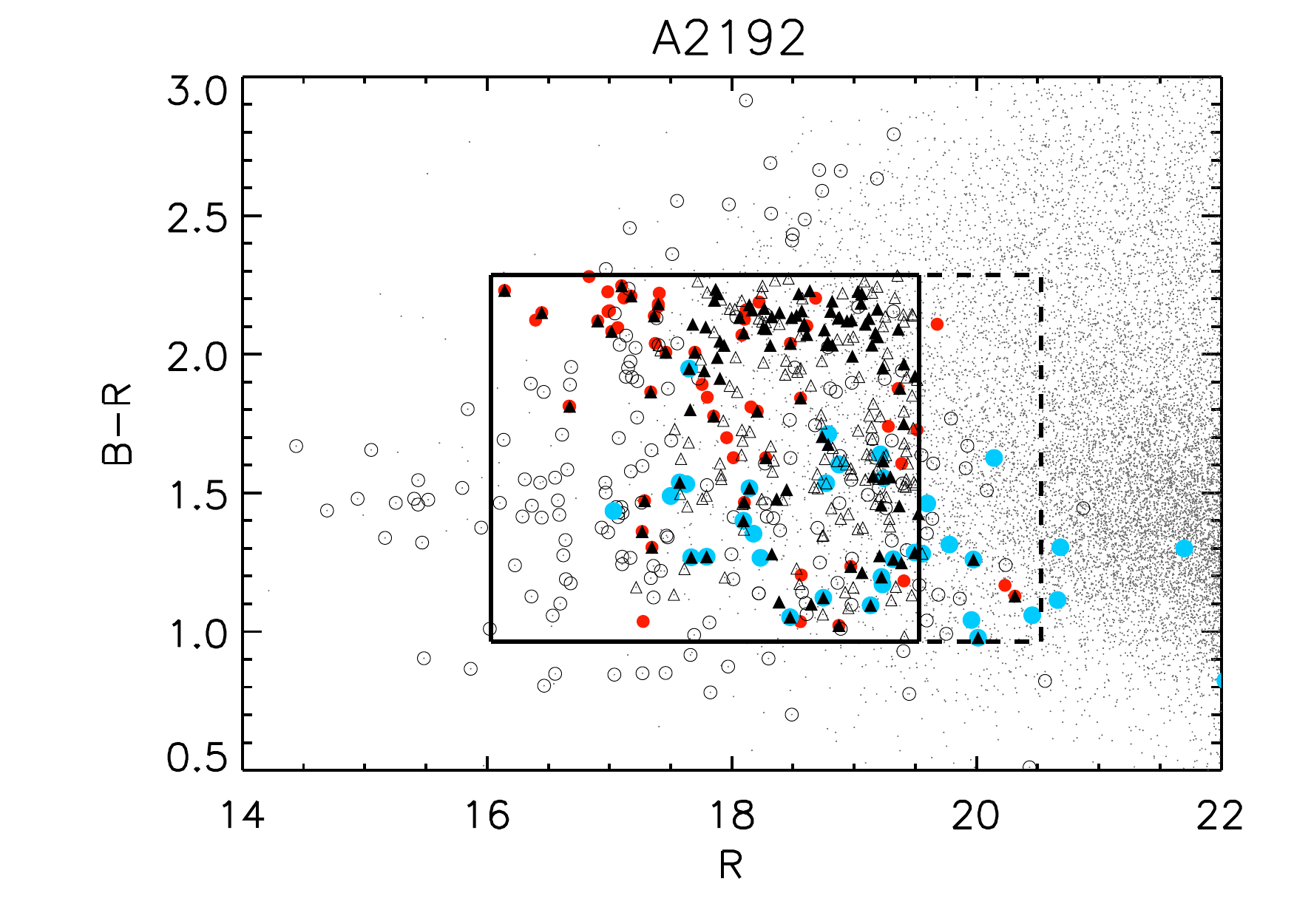}
\end{center} 
\caption{The CMDs of A963 (left) and A2192 (right). 
All galaxies in the SDSS photometric catalogs and in our INT images are plotted in small gray dots. The red filled circles indicate galaxies with literature redshifts that lie inside the HI redshift range ($0.164\leqslant z \leqslant0.224$), whilst the black open circles are those with literature redshifts outside this range. The bigger blue filled circles highlight the HI-detected galaxies. Solid triangles correspond to new  WHT redshifts inside the HI redshift range, and open triangles are those with WHT redshifts outside the range. In each case, the solid box delimits the region containing the galaxies targeted for spectroscopy, and the extended dashed area shows the fainter range in which only galaxies with HI detections were targeted.}
 \label{CMDs}
\end{figure*}

\textbf{(ii) Location in the colour-magnitude diagram} (CMD): For each field we created CMDs showing $R$ versus $B-R$ (see Figure~\ref{CMDs}) for all the sources detected in the photometry that were classified as galaxies (c.f. bullet point (i)). 
We selected galaxies within the black solid boxes shown in the CMDs of Figure~\ref{CMDs} and included fainter HI-detected galaxies (up to $R=20.5$). This was done partially to confirm clear optical counterparts of fainter HI-detections.  The magnitude limits for each field are listed in Table ~\ref{maglim_table}.  Figure~\ref{CMDs} shows that the red sequence is well-defined in both clusters. The figure also shows that the vast majority of HI-detected galaxies are located in the so-called ``blue cloud''.  

\begin{figure*}
\begin{center}
  \includegraphics[width=0.9\textwidth]{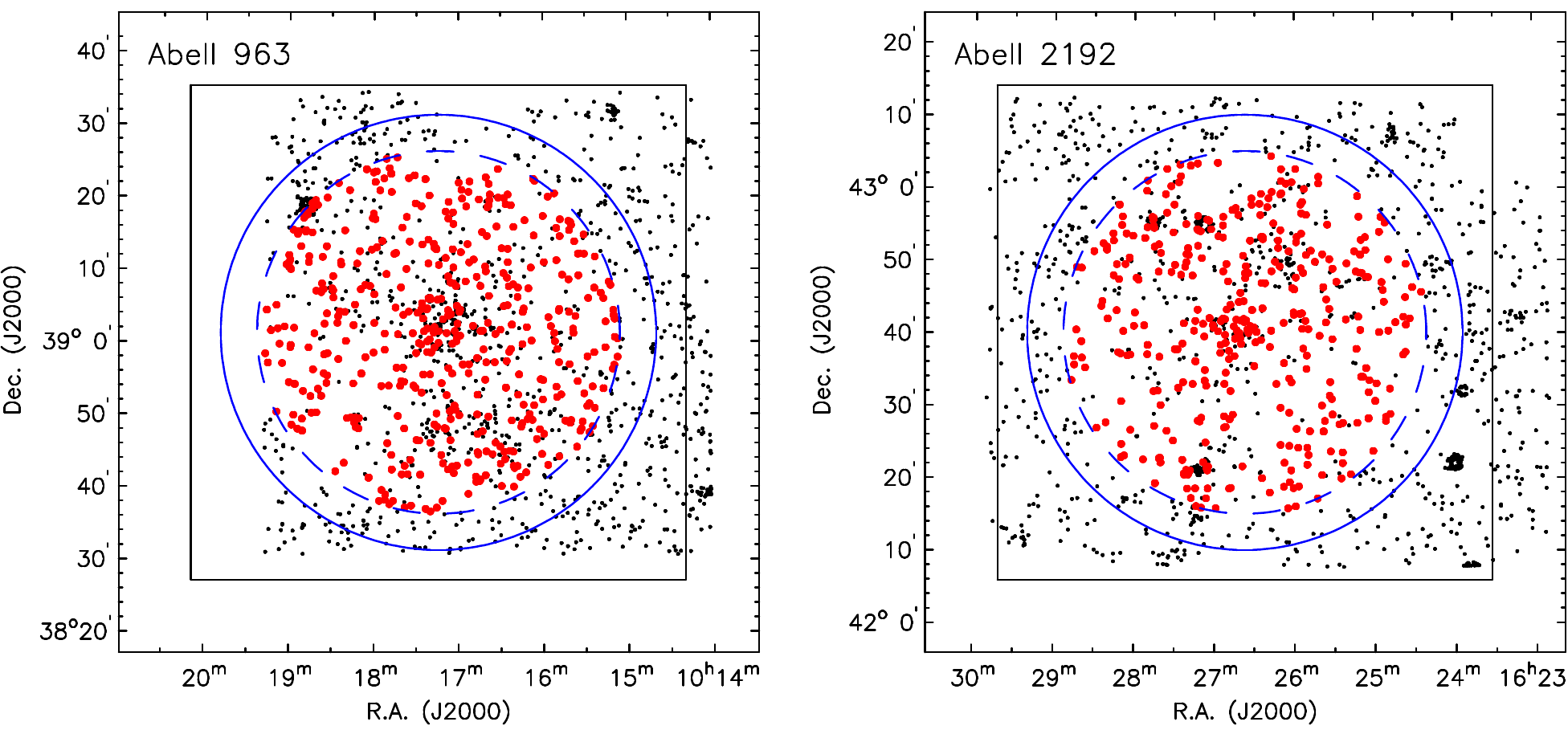}
\end{center} 
\caption{The distribution of targeted  (black points) and observed (red filled circles) galaxies in A963 (left) and A2192 (right). The black square represents the area within $68\times 68$ arcmin$^2$ of the WSRT pointings. The bigger circle (30' radius) encloses the field of view of WHT, whilst the smaller dashed circle (25' radius) shows the unvignetted area.}
 \label{FOVs}
\end{figure*}

\textbf{(iii) Area in the sky:} Although fibers can be positioned over a field of 1 degree diameter, we only selected galaxies inside a radius of 25 arcmin (centered on the pointings of the HI observations), as it is the radius within which vignetting is not dramatically strong.  The spatial distribution of the targeted galaxies (red circles) and the field of view of the WSRT and WHT are shown in Figure~\ref{FOVs}.

\textbf{(iv) Prioritizing the fiber allocation:} When allocating fibers to objects, we gave the top priority to HI-detected galaxies, lower priority to other targeted galaxies, and the lowest priority to galaxies that already had redshifts from the literature\footnote{Whenever we refer to ``literature redshifts'' we are referring to SDSS, WYIN, Larvery \& Henry and Czoske (see Section~\ref{sec:data}).} in the range of the HI survey (see Table~\ref{targ_table}). Furthermore, when the literature redshifts were outside the cluster redshift range (open black circles in Figure~\ref{CMDs}), we explicitly rejected the galaxies.

\begin{table}
\begin{center}
 \caption{For each field the table lists: the number of galaxies with redshifts available from the literature in the two surveyed fields (within $0.164\leqslant z \leqslant0.224$), number of HI-detected galaxies,  number of 
galaxies targeted for WHT spectroscopy (fullfilling the criteria listed in bullet-points i to iii in Section~\ref{subsec:target}), and number of new redshifts obtained in our spectroscopic campaign (with quality flag $\geq-1$, c.f. Section~\ref{subsec:redshifts}).} 
\label{targ_table}
\begin{tabular}{cccccc}
\hline\\[-1mm]
Field 	& No. lit. z 	& No. HI 	 &No. WHT 	& No. WHT z	\\
	& in HI-range 	& galaxies	 &targets 	& in HI-range		\\
\hline\\[-2mm]
A963	&	161	&	119	& 853		&	261	\\
A2192	&	67	&	37	& 612		&	251	\\
\hline\\
\end{tabular} 
\\ 
\end{center}
\end{table}

\subsection{Observations and data reduction}
\label{subsec:obs}

\begin{figure*}
\begin{center}
  \includegraphics[width=0.49\textwidth]{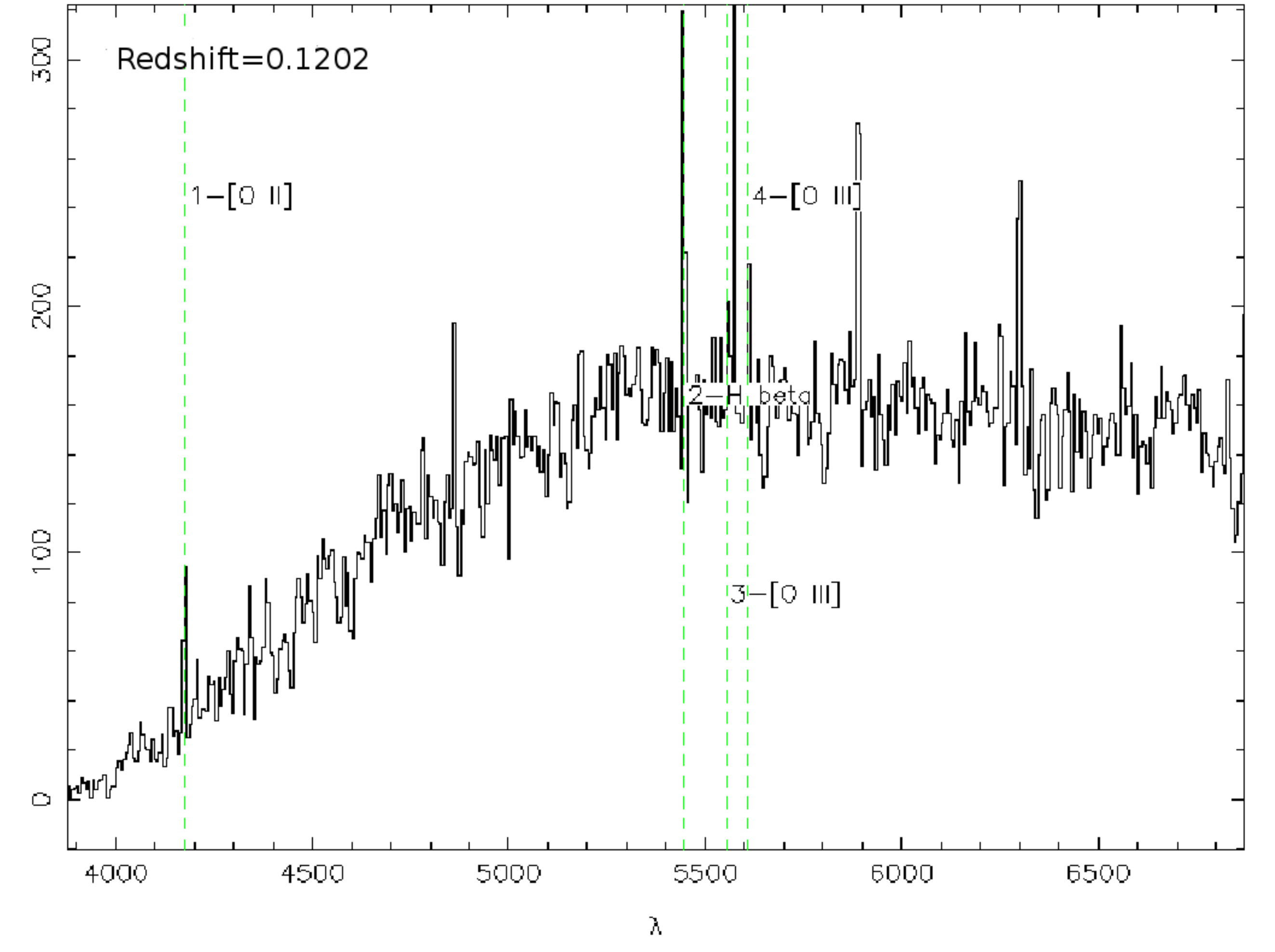}
  \includegraphics[width=0.49\textwidth]{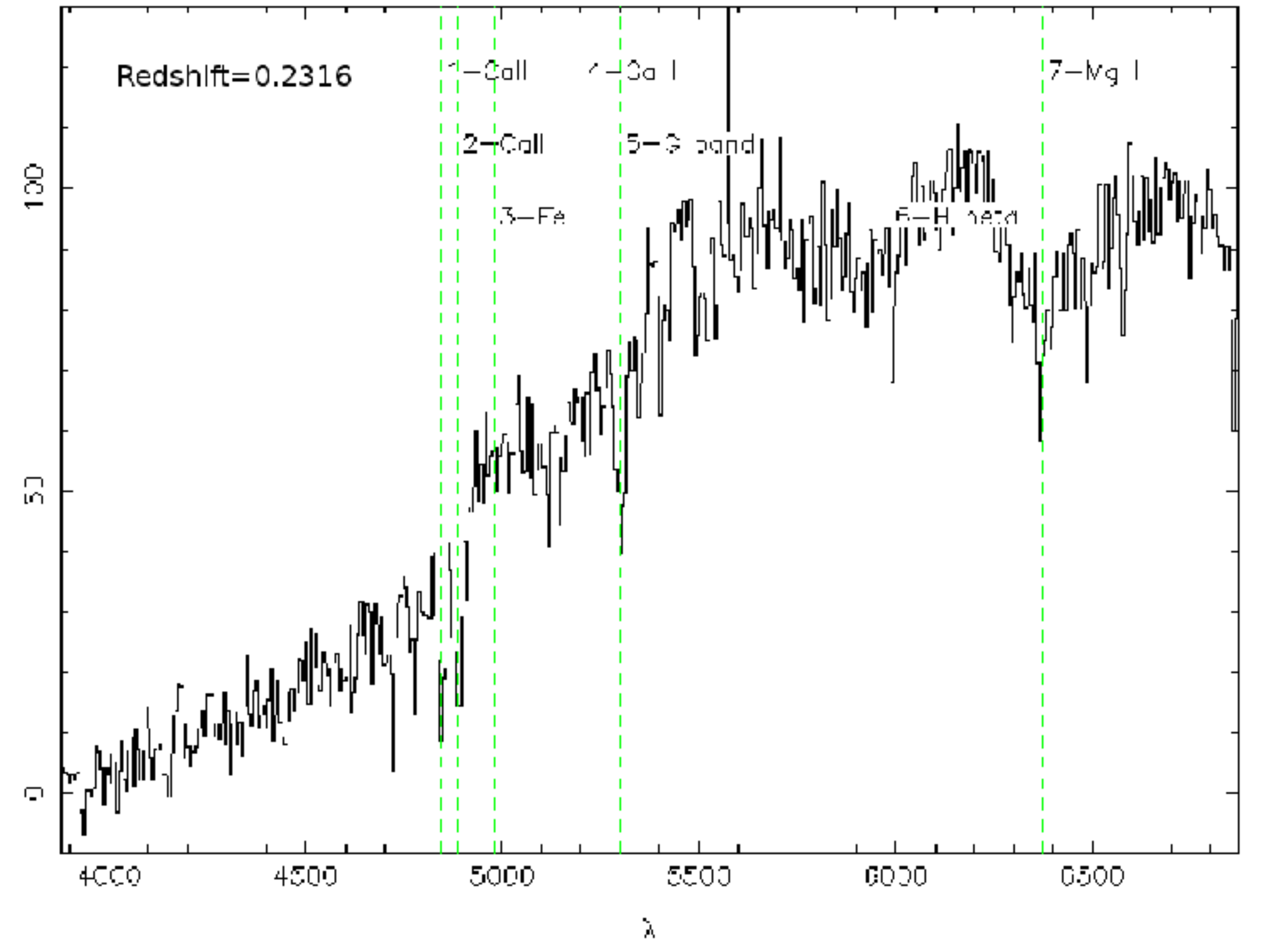}
\end{center} 
\caption{Example of the extracted, calibrated and sky-subtracted spectra for a galaxy with strong [OII] emission, as well as some absorption features (left) and an absorption-line galaxy (right). The wavelength  ($\lambda$) is expressed in $\rm \AA$ and the flux (vertical axes) in arbitrary units. Green vertical lines indicate the main lines (such as [OII] emission or the Ca H\&K absorption lines) from which the redshift was calculated. The measured redshift in each case is also indicated at the top left corner of each spectra.} 
 \label{speceg}
\end{figure*}

\begin{table*}
\begin{center}
 \caption{Magnitude and colour limits for spectroscopic target selection. The colours cuts were applied to all samples, and the magnitude limits are those highlighted with boldface (see also the red boxes in Figure~\ref{CMDs}). We split the galaxy sample into ``bright'', ``faint'' and ``fainter HI-galaxies'', to produce configurations of the same exposure time, and maintain the signal-to-noise above 10. The bright configurations were observed for $\geq$1 hour, whilst the faint configurations (including the fainter HI-galaxies) were observed for $\geq$3 hours. }
\label{maglim_table}
\begin{tabular}{ccccc}
\hline\\[-1mm]
Field 		&  Bright    		& Faint		& Fainter  	& Colour	 \\
		 & sample 		& sample	& HI-galaxies 	& constraints	 \\
\hline\\[-2mm]
Abell 963    	&\textbf{15.0} $< R < 18.5$	& $18.5< R <$ \textbf{19.5}	&$19.5<R<21.0	$	&$0.98<B-R<2.38$	\\
Abell 2192	&\textbf{16.0} $< R < 18.5$	& $18.5< R <$ \textbf{19.5}	&$19.5<R<20.5	$	&$0.92<B-R<2.24$	\\
\hline\\
\end{tabular} 
\\ 
\end{center}
\end{table*}

The spectroscopic observations were made using the AutoFib2+WYFFOS (AF2) wide-field, multi-fiber spectrograph mounted on the 4.2m  WHT in La Palma.
AF2 contains 150 science fibers (of 1.6 arcsec diameter and 26 metres in length), and 10 fiducial bundles for acquisition and guiding. At the prime focus, the fibers are placed onto a field plate by a robot positioner at user-defined sky coordinates. Object light is transmitted along the fibers to the spectrograph. 

We used the R600B grating with the 2-chip EEV 4300$\times$4300 mosaic that counts with 13.5$\mu$ pixels. Using a 2$\times$3 binning of the CCD pixels, we obtained a spectral resolution of $\sim$4$\rm \AA$ FWHM (depending on the location on the CCD). The spectra were centered on a wavelength of  $\sim4900 \rm \AA$ and covered the range $\sim 3900-6900 \rm \AA$. In this range, galaxies at the targeted redshift showed spectral features from the Ca H\&K lines, and the [OII] line in the blue, up to NaD in the red (see Figure~\ref{speceg}).  We used He and Ne lamp exposures for wavelength calibration.

We obtained the data in 2 runs (April and June 2011) during 5 dark nights in total. In this time, we were able to observe 12 fiber configurations, each containing $\sim 80$ galaxies. The targeted galaxies were divided into different configurations (see Table~\ref{maglim_table}) depending on their luminosities: for each field we created 3 bright and 3 faint configurations. For the bright ones, the exposure times were chosen to be one hour and for the faint ones 3 hours, in order to recover a signal-to-noise $\gtrsim 10$, although in practice, we extended the exposure times when possible. We tried to maximize the number of fibers allocated on galaxies but also placed typically 20-30 fibers on the sky, for sky subtraction purposes.

As explained in Section~\ref{subsec:target}, we prioritized our targets according to their HI content, and previous observations. 
Although we did not target the same galaxy more than once, in some cases, the software that allocates fibers to coordinates in the sky  (\begin{small}AF2\_CONFIGURE\end{small}) used fibers, that could not be allocated to a new galaxy, to target galaxies previously observed in another configuration. For this reason, we have a few dozen galaxies with repeated observations (in different configurations, and sometimes in different runs), which were helpful for characterizing our redshift errors and the quality of our data.

\begin{figure*}
\begin{center}
  \includegraphics[width=0.49\textwidth]{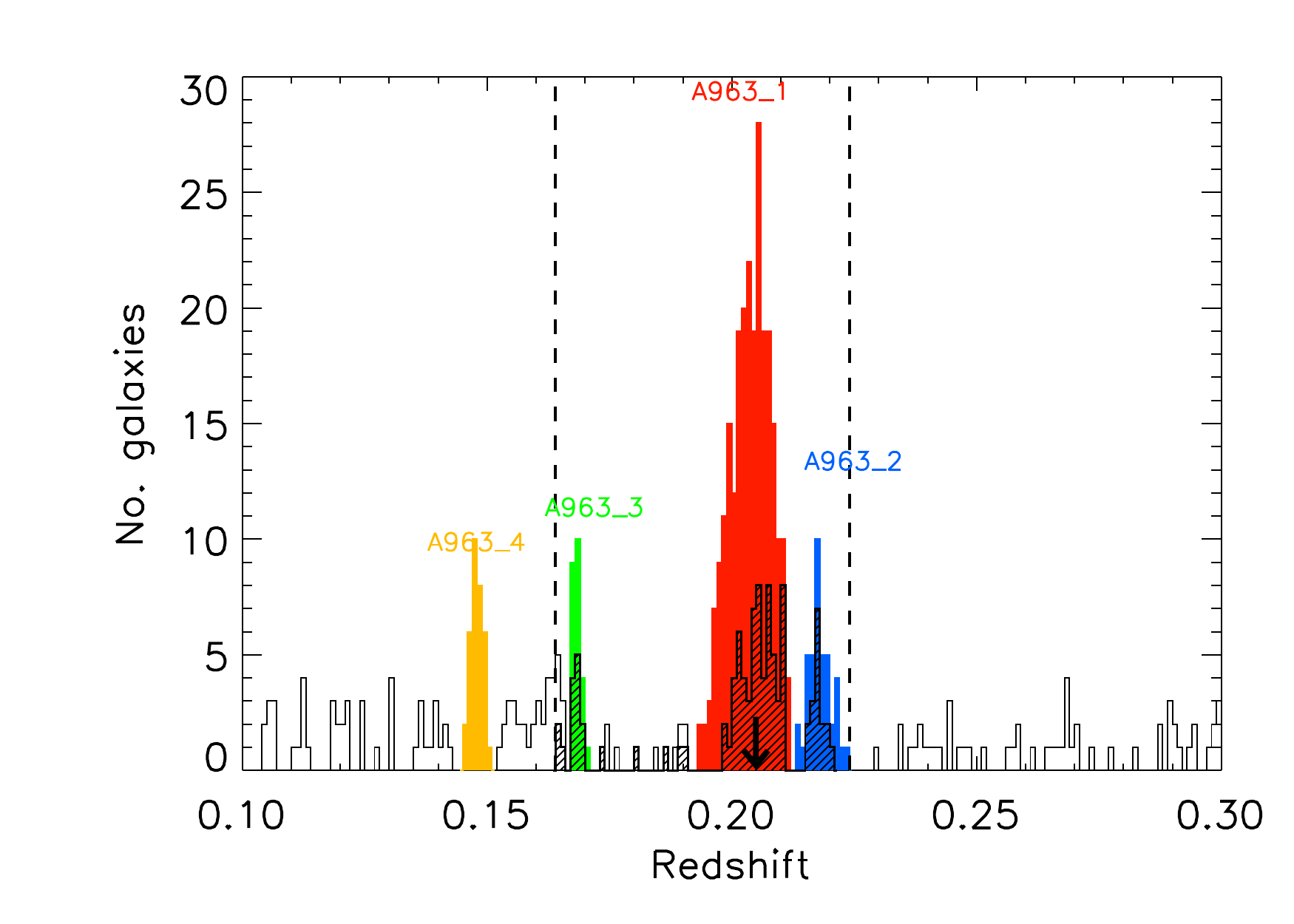}
  \includegraphics[width=0.49\textwidth]{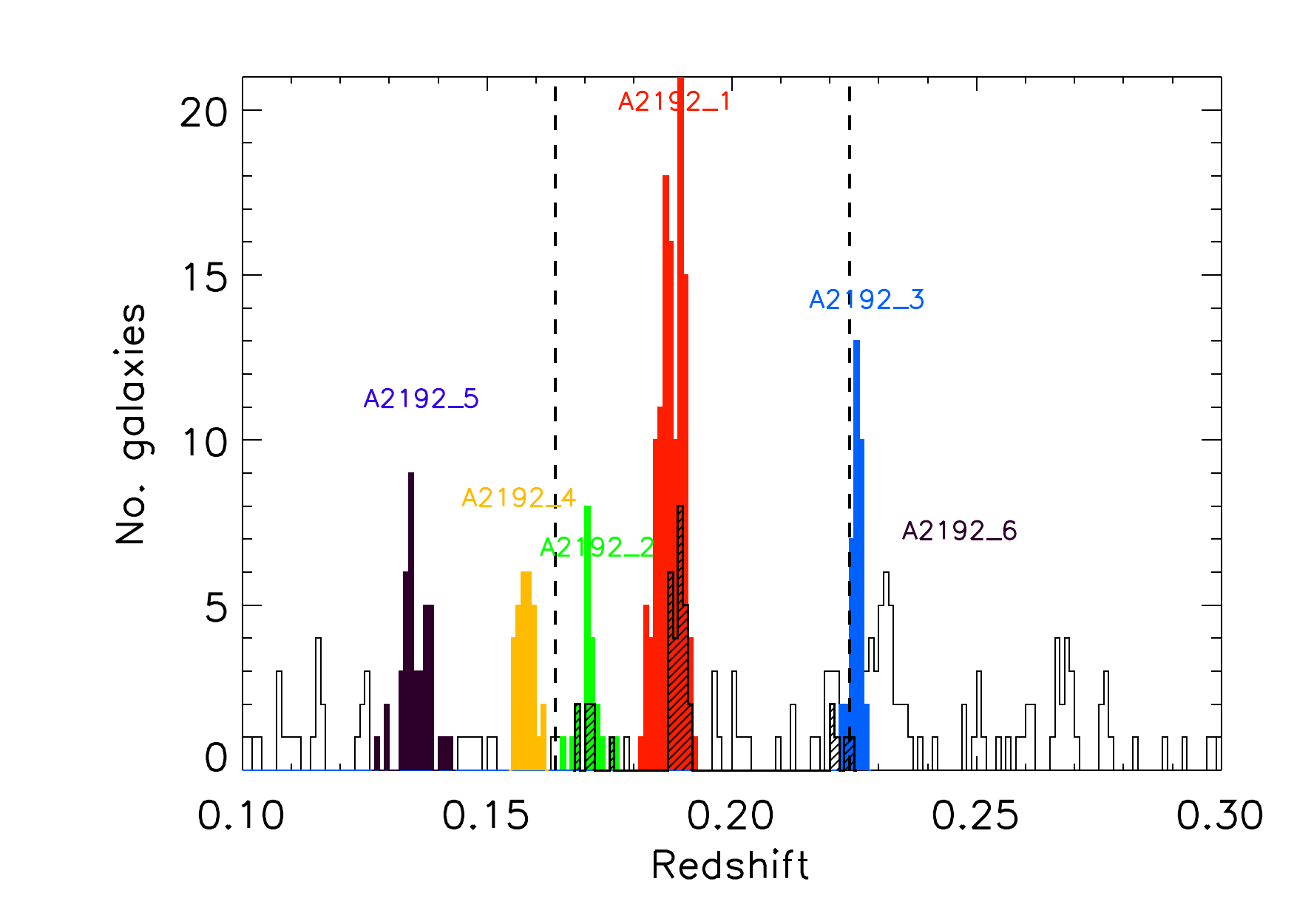}
  \includegraphics[width=0.49\textwidth]{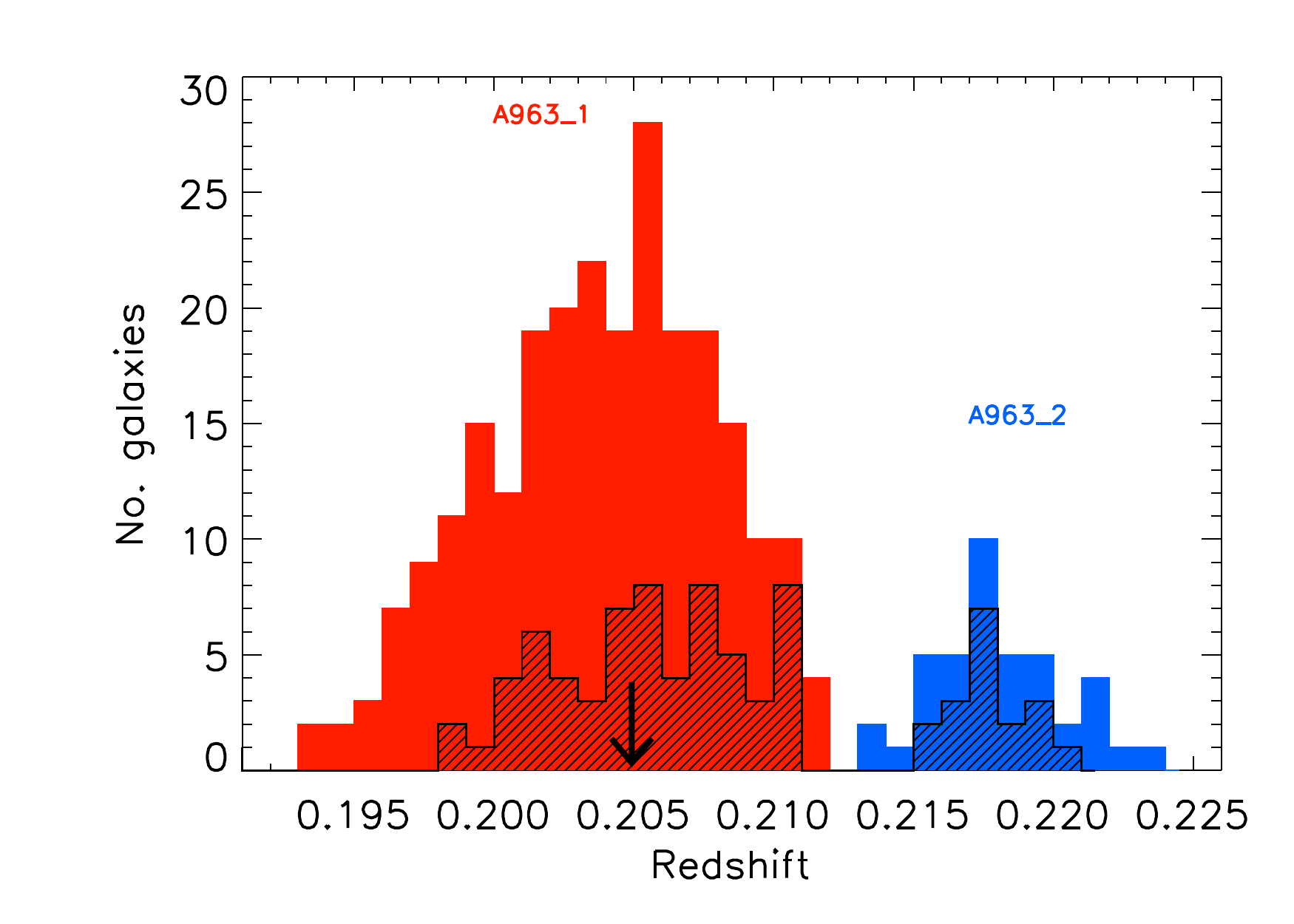}
  \includegraphics[width=0.49\textwidth]{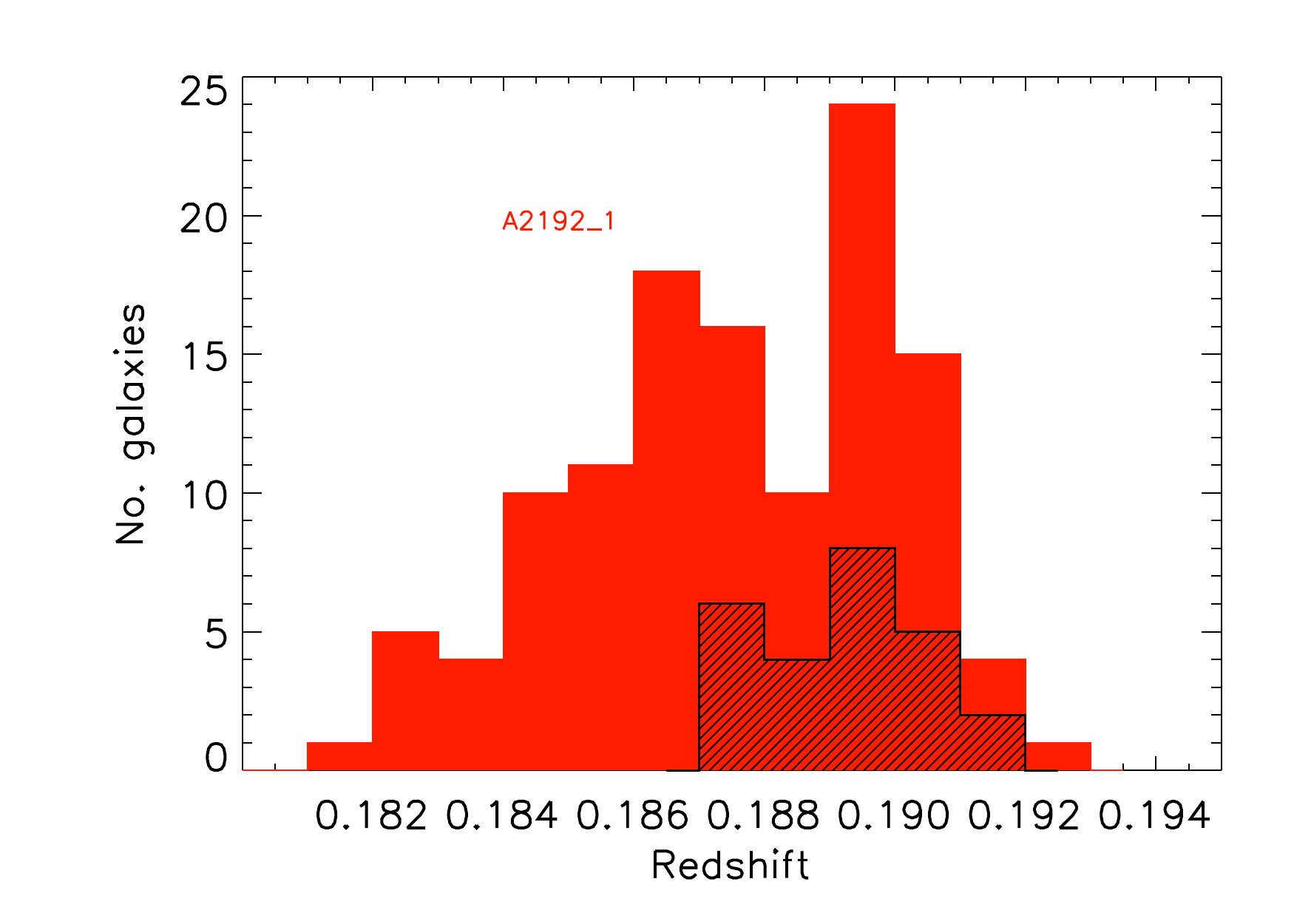}
  \includegraphics[width=0.49\textwidth]{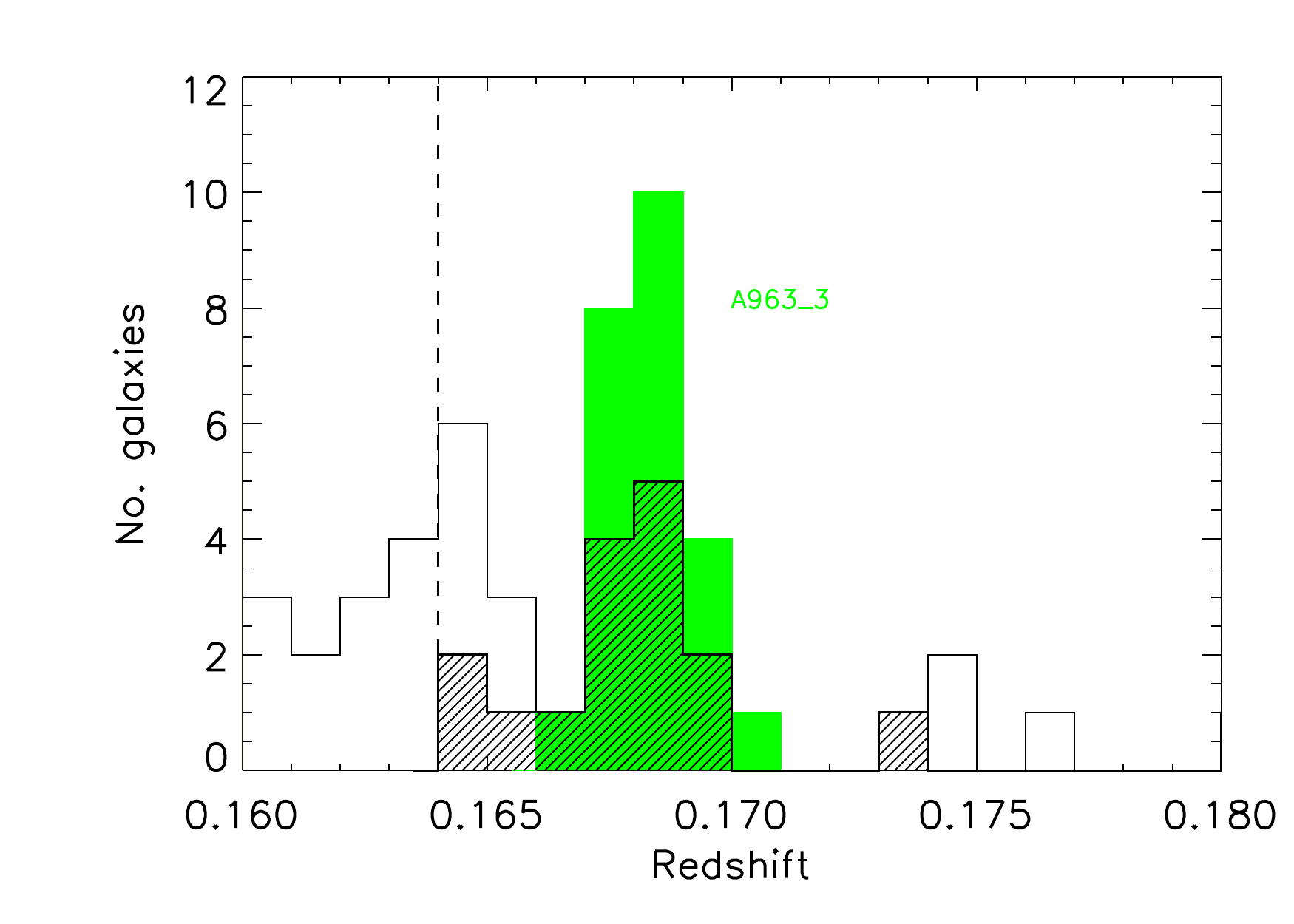}
  \includegraphics[width=0.49\textwidth]{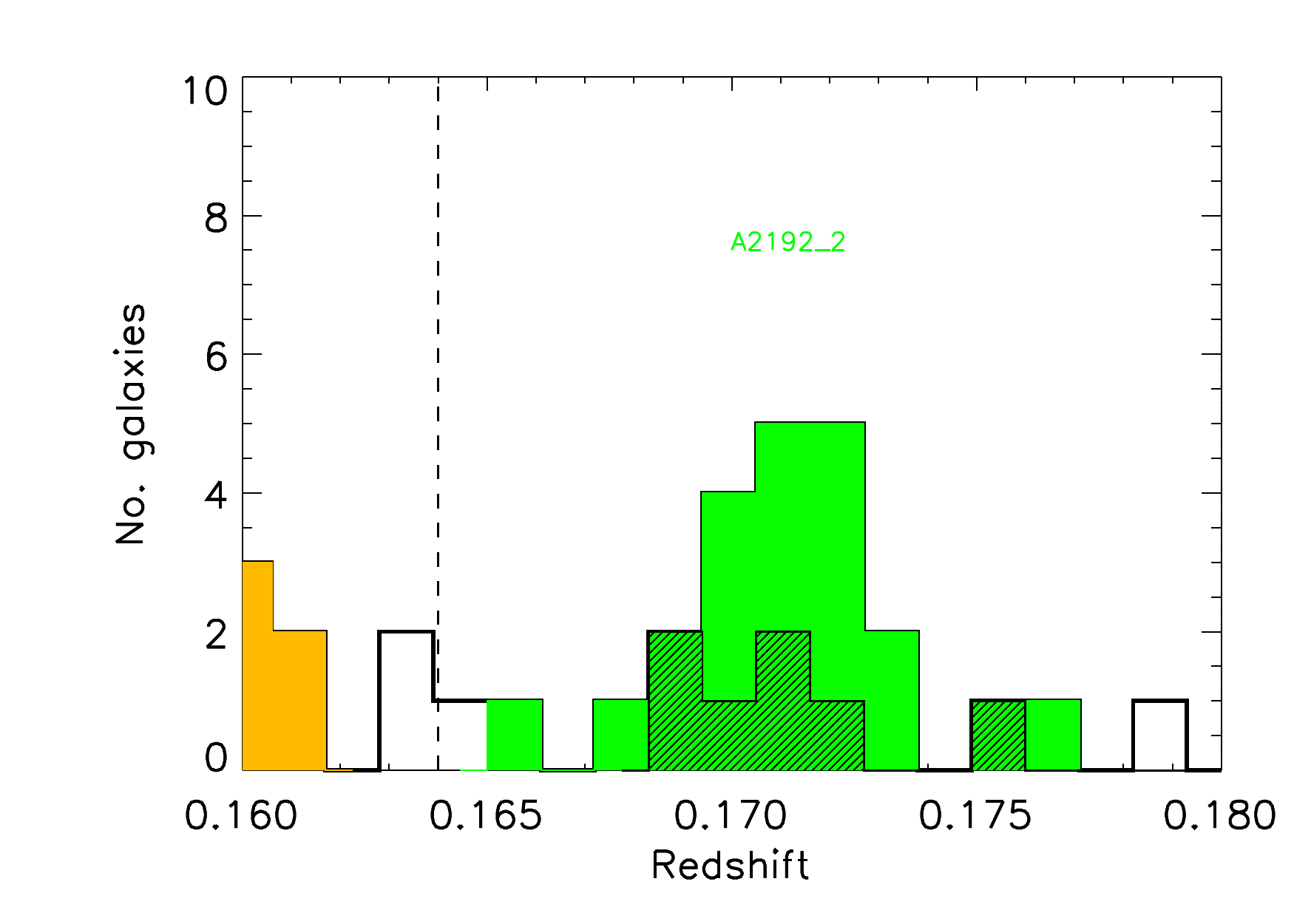}

\end{center} 
\caption{Redshift histograms for A963 (left) and A2192 (right). The top panels show the full redshift range of BUDHIES (The limits of the HI survey are indicated with vertical dashed lines), the middle panels a zoom in on the main clusters, and bottom panels additional groups.  The open histograms correspond to the distribution of all galaxies with a redshift (of all qualities). The filled-dashed histograms show the HI-detected galaxies for reference, although we note that thesw will be fully analysed in a subsequent paper. 
The different structures identified are represented in different colours, as described by the labels. In the case of A963\_1, the redshift of its brightest cluster galaxy is indicated with an arrow.   
%The vertical dashed lines delimits the redshift range of the HI survey,
%. The middle panels of the figure show zoomed areas of the redshift distributions on the top centered on the main cluster(s), while the bottom panels show additional groups.
}
 \label{zhist}
\end{figure*}

The data were reduced using the new AF2 data reduction pipeline\footnote{For description and  download of the pipeline visit: 
http://www.ing.iac.es/Astronomy/instruments/af2/pipeline.html}. We started using the pipeline before the final version was released, and hence helped testing the pipeline. Our final spectra however were reduced with the latest (final) version. To control the quality of our final reduction, we compared the final product with (a subsample of) the same spectra reduced with IRAF (using the IRAF package \textit{dofiber}), coming to the conclusion that the pipeline performs well, and that it is significantly faster and more efficient than \textit{dofiber}.

The pipeline is written in IDL and is able to perform full data reduction, including fiber to fiber sensitivity corrections and optimal extraction of the individual spectra. The first steps of the pipeline include master bias correction, tracing of the fibers, flat-field correction, and masking of bad pixels in the science data. Twilight sky flats were used to define the apertures and trace the spectra on the CCD, and to perform the flat-field correction. An optimal extraction algorithm is used for extracting the spectra.  The software accounts for wing emission from adjacent fibers.

As for the wavelength calibration, the software selects an arc lamp spectrum from a fiber near the centre of the chip. The user is then prompted to identify the principle arc lines (between 10 and 12 in our case). Other significant lines are automatically identified and then a dispersion solution is found using a polynomial fit. Once the user is satisfied, the dispersion solution is propagated to the rest of the fibers. 
Typically, the fits yielded an rms scatter of 0.03 $\rm \AA$.

Because not all fibers have the same throughput, we scaled the final spectra according to an estimation of the throughput in each fiber that the pipeline estimates from the master flat.

A master  sky spectrum was also derived for each exposure by combining the spectra (using the median) of the 20-30 individual fibers assigned to the sky. 
The median is then subtracted from each science spectrum. 

Finally, the multiple exposures in each pointing were combined (after extraction) with cosmic rays being removed.

Figure~\ref{speceg} shows example reduced spectra, and 
Table~\ref{targ_table} summarizes the target selection and the outcome of our spectroscopic campaign.

\begin{figure*}
\begin{center}
  \includegraphics[width=0.49\textwidth]{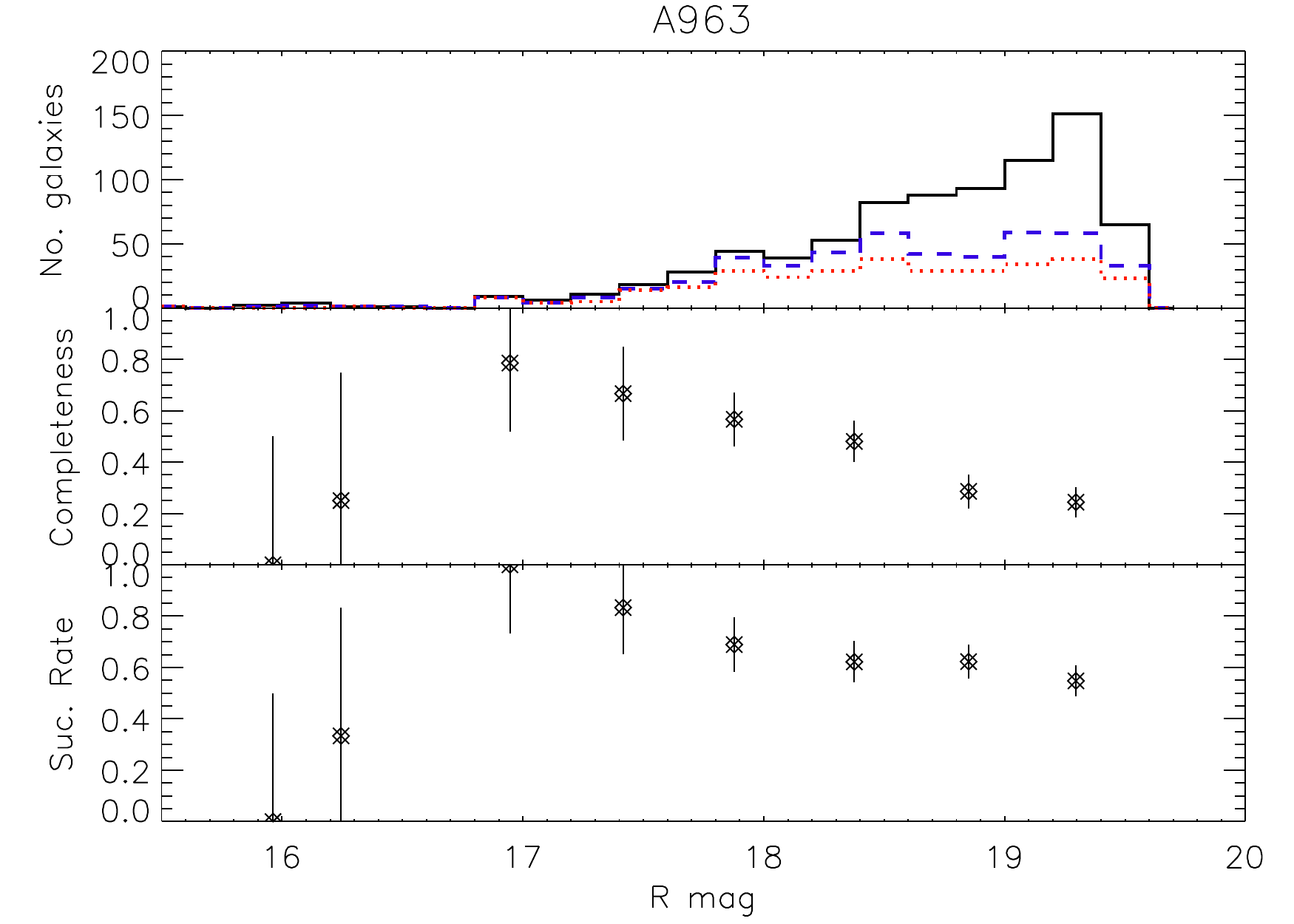}
  \includegraphics[width=0.49\textwidth]{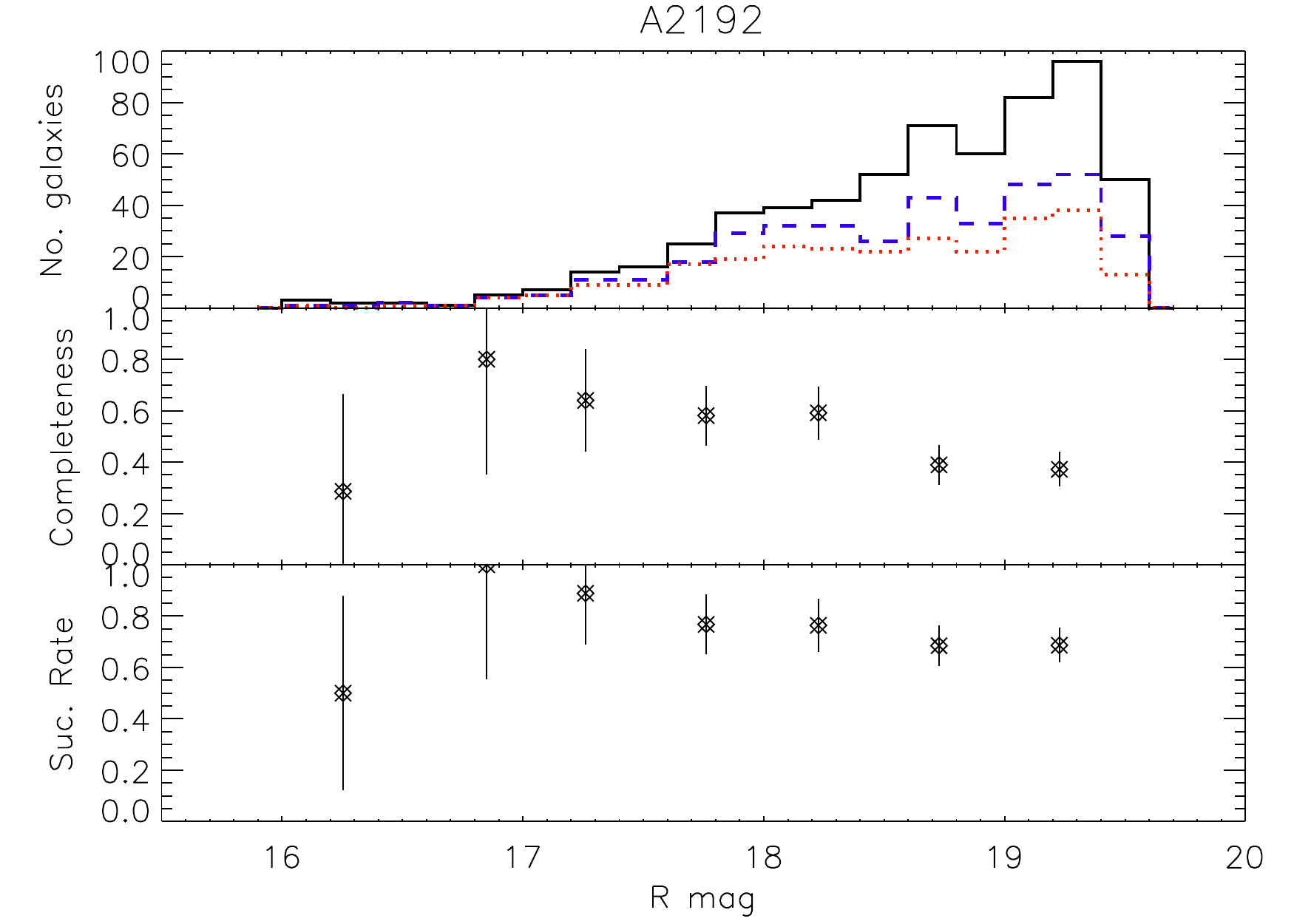}
\end{center} 
\caption{The top panels of this figure show the magnitude distribution of the targeted sample (solid histogram), in addition to the observed targets (blue dashed histogram) and the observed targets with trusted redshifts (red dotted histogram) for A963 (left) and A2192 (right). The middle and bottom panels show the completeness and success rate (respectively) as a function of magnitude. All the available redshifts were used to compute the completeness. }
 \label{compl_hist}
\end{figure*}

\subsection{Galaxy Redshifts and EW[OII]}
\label{subsec:redshifts}

Spectroscopic galaxy redshifts were measured using emission lines where possible (typically the [OII]λ3727$\rm \AA{}$ line), or the most prominent absorption lines (e.g. Ca H\&K lines at 3934$\rm \AA{}$ and 3968$\rm \AA{}$), as shown in Figure~\ref{speceg}. The redshifts were manually assigned a quality flag of 1 (reliable redshifts), -1 (reliable redshifts but with larger uncertainties) or -2 (unreliable redshifts). Over half (60\%) of the measured redshifts are of high quality ($\geqq-1$). Some of the objects however, have an unreliable redshift (10\%), or in some cases no redshift could be determined (30\%). 
% The complete version of the table is electronic format only). 
A table containing all the measured redshifts (and EW[OII]) is available in the electronic version of the paper (see Table C1 for a shorter version).

In total, we measured 512 new reliable (quality $\geq-1$) redshifts in A2192 and A963. The redshift distributions for the surveyed volumes are shown in the top panels of Figure~\ref{zhist}.  
Note that the redshift distribution of the HI-detected galaxies is overplotted in the histograms for reference, although the complete analysis of the distribution and properties of the HI-detected galaxies  will be presented in subsequent papers. 
We refer to \citet{Jaffe2012} for first results on the HI distribution in A2192's main cluster. 
%First results on the HI distribution in A2192's main cluster can be found in \citet{Jaffe2012}. 

We estimate the typical redshift error from galaxies that have been observed more than once (i.e. in more than one configuration) and we get an uncertainty of $\backsimeq$0.0003. 
We further cross-checked our redshift measurements with previous spectroscopic observations, available for a substantial number of galaxies in our sample and confirmed the quality of our redshift measurements.

%\subsection{EW[OII]}

In addition to the redshifts, we also measured rest-frame equivalent widths (EW) of the [OII]3727$\rm\AA{}$ line, 
which will be used as a proxy for ongoing star formation in forthcoming papers. 
The EW[OII] values are  listed in Table~\ref{spec_table} for future reference.

\subsection{Completeness and success rate}
\label{subsec:completeness}

For our analysis in this and other forthcoming papers, it is very important to know the spectroscopic completeness, as it will affect local densities, or any magnitude-dependent analysis (e.g. the different galaxy population fractions inside clusters).

First, we assessed the completeness as a function of magnitude ($m$) as follows:

\begin{equation} 
\label{completeness}
 C(m)=\dfrac{N_z}{N_{\rm tar}(m)}
\end{equation}

\noindent where, $N_{z}$ is the number of galaxies with reliable redshift and $N_{\rm tar}$ is the total number of targeted galaxies for spectroscopy (i.e. all the galaxies in our photometry that lie inside the CMD box and are inside the unvignetted field-of-view of the WHT, c.f. Section~\ref{sec:spec}).

The success rate, i.e. the fraction of galaxies with trusted redshift determination with respect to the total number of galaxies observed, is defined as:

\begin{equation}
\label{SR}
 SR(m)=\dfrac{N_z}{N_{\rm obs}}(m)
\end{equation}

\noindent where, $N_{obs}$ is the number of galaxies spectroscopically observed.

This assessment of the completeness only gives a general idea of our completeness limits, as shown in Figure~\ref{compl_hist}. In the upper panel,  the magnitude distribution of the targeted, observed and reliable-redshift samples are shown, and in the lower panels  the completeness and success rate as a function of magnitude are plotted.
This plot shows that our sample is complete to $>40$\% almost across all magnitudes.

To accurately correct for completeness, we further quantified possible changes with colour and spatial distribution. We did this by computing the completeness in colour-magnitude bins and in $\alpha-\delta$ bins. Colour bins were $\sim$0.5 magnitudes wide, and geometrical bins were $\sim$0.3$^{\circ}$. The $R$-band magnitude bins were larger towards the brighter end to allow a similar number of galaxies in each colour-magnitude bin. The geometrical effects could be caused by fiber collisions in the cluster centre, where there is more crowding. The effect is expected to be small as we had several configurations of the same cluster. Nevertheless, a geometrical completeness (C$_{\rm geo}$) was calculated after applying the colour-magnitude completeness correction (C$_{\rm CM}$).

The colour-magnitude and geometrical completeness functions will be applied to the spectroscopic sample in the following papers of the series to calculate fractions of a given galaxy population (e.g. the fraction of HI-detected galaxies as a function of environment).  
This can be achieved by weighting each galaxy by $C_{\rm CM} + C_{\rm geo}$, where $\Sigma (C_{CM}) = N_{\rm tar}$ and $\Sigma (C_{\rm geo}) = 1$.

\begin{figure}
\begin{center}
  \includegraphics[width=0.49\textwidth]{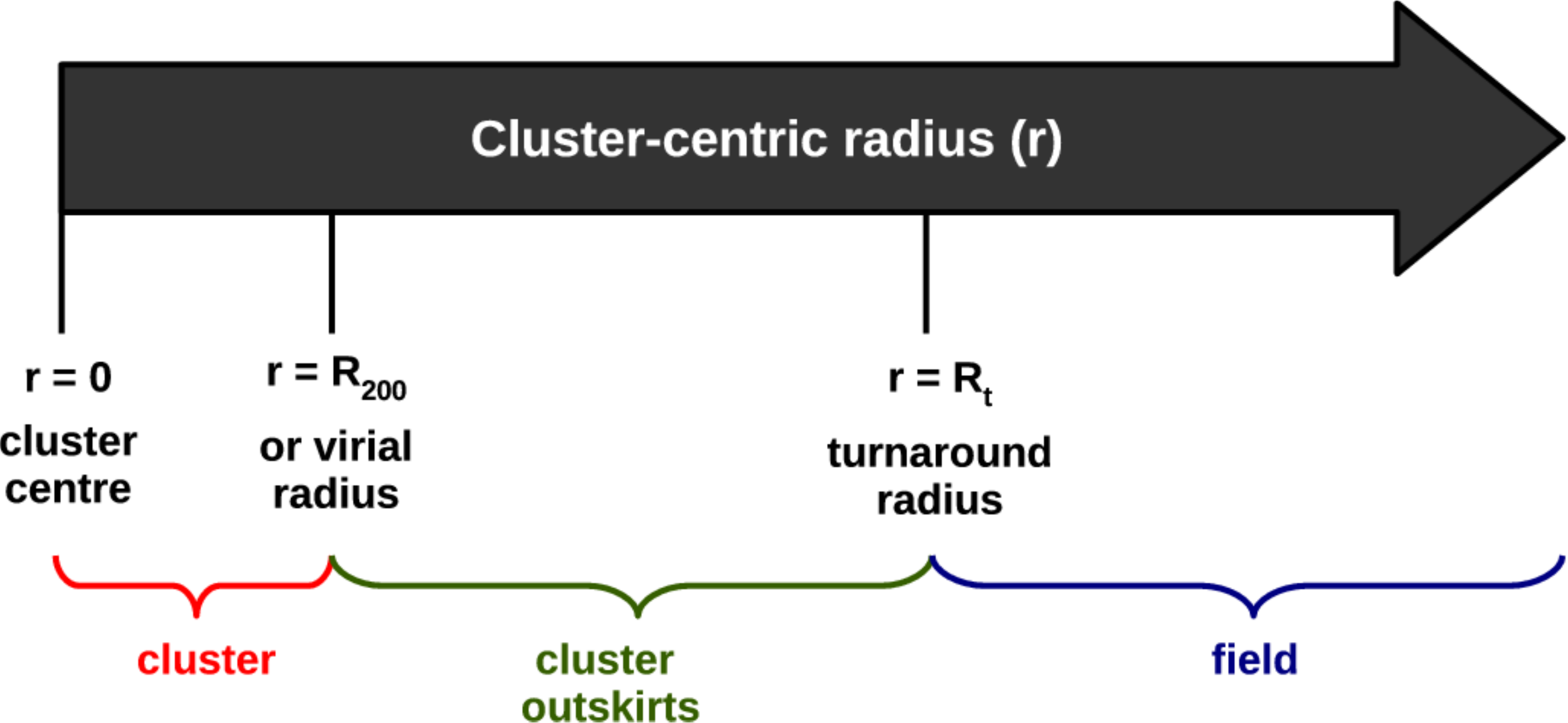}
\end{center} 
\caption{Illustrative schema of our environment definition criteria. The arrow represents the cluster-centric radius, going from the cluster's central region ($0<r<R_{200}$), passing through the cluster outskirts ($R_{200}<r<R_{\rm t}$), and ending in the field ($r>R_{\rm t}$). }
 \label{schema1}
\end{figure}

\section{velocity dispersions and cluster/group membership}
\label{sec:mem}

To identify the main clusters and other structures in the fields of A963 and A2192, we used the following approach. 

Using galaxies with high quality redshifts ($\geq-1$), we constructed redshift histograms and identified (by eye) possible galaxy overdensities. This is shown in the top panels of Figure~\ref{zhist}, where all the identified structures are highlighted.  

Following the approach of \citet*{bfg90},  we computed the cluster velocity dispersion and central redshift in each case. 
Equation~\ref{rfv} shows the definition of the peculiar velocity of a galaxy with redshift z in the rest frame of a cluster with redshift z$_{\rm cl}$. This equation is valid to first order for $v \ll c$ \citep{Harrison1974,Carlberg1996}. The velocity dispersion of the cluster ($\sigma_{\rm cl}$) is then defined as the dispersion of the $v$ values for the cluster members. This value is related to the observed velocity dispersion ($\sigma_{\rm obs}$), as shown in Equation~\ref{veldisp}.

\begin{equation}
\label{rfv}
 v=c\dfrac{z-z_{\rm cl}}{1+z_{\rm cl}}
\end{equation} 

\begin{equation}
\label{veldisp}
 \sigma_{\rm cl}=\dfrac{\sigma_{\rm obs}}{1+z_{\rm cl}}
\end{equation}

Because the velocity dispersion of a cluster is a proxy for cluster mass \citep[see][]{Finn2005}, we use it to distinguish galaxy clusters from less massive groups. Although there is no strict velocity dispersion cutoff for separating groups from poor clusters, in our environmental definition, we adopt a threshold value of $500$km s$^{-1}$, following \citet{Mulchaey2000}.  
%consider \textit{clusters} the structures with  $\sigma_{\rm cl}>500$km s$^{-1}$ and \textit{groups} those with $\sigma_{\rm cl}<500$km s$^{-1}$ 

We further calculated $R_{200}$, the radius delimiting a sphere with a mean density equal to 200 times the critical density\footnote{$R_{200}$ is commonly used as an equivalent of virial radius.}, as in \citet{poggianti2006}:

\begin{equation}
\label{r200}
R_{200}=1.73\dfrac{ \sigma_{\rm cl}}{1000 \rm km \rm s^{-1}}\dfrac{1}{\sqrt{\Omega_{\Lambda} +  \Omega_{0}(1+z_{\rm cl})^3 }} h^{-1} Mpc
\end{equation} 

To carefully distinguish between environments, we classify galaxies within or around a cluster or group in the following categories: cluster, cluster outskirts and field.  
For this, we make use of $R_{200}$ and the so-called turnaround radius, $R_{\rm t}$, that separates the infall region and the field. Following the work of \citet{RinesDiaferio2006}, we assume that $R_{\rm t}$/$R_{200}=4.57$.
We define \textit{cluster} galaxies to be inside  $R_{200}$, \textit{outskirt} galaxies to lie between  $R_{200}$ and $R_{\rm t}$, and \textit{field} galaxies to be beyond $R_{\rm t}$, as shown schematically in Figure~\ref{schema1}. 

Table~\ref{structures} lists the $\sigma_{\rm cl}$, $R_{200}$, central redshift and number of members for each structure found inside the HI survey's volume. Table~\ref{structures_2} lists other structures identified outside the studied redshift range.

%\movieref[3Dgetview]{pt}{Click here!}
%\movieref[3Dgetview]{pt2}{Click here!}

\begin{figure*}
\begin{center}
\includemovie[
  3Dc2c=-0.9436892867088318 -0.31196436285972595 0.11013055592775345, 
  3Droo=6.363637597521383, 
  3Droll=-9.655947712673747,
  3Dbg=1. 1. 1.,
  3Dlights=CAD,
  3Drender=Solid,
    3Dviews2=movs/views_963.tex,
%    3Droo=7,3Dlights=CAD,
    poster=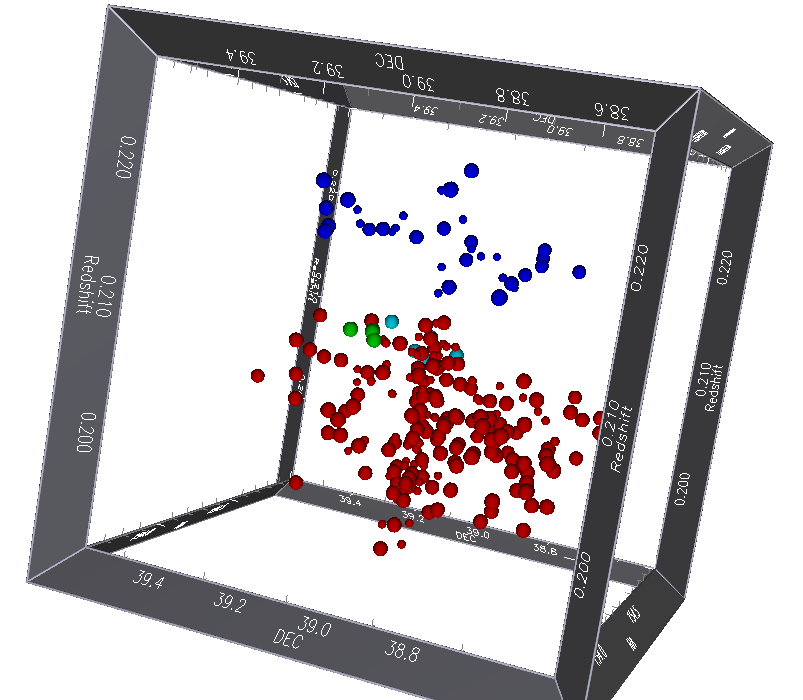,
    toolbar,
    label=pt,
%    %text={click to ativate},     
    ]{0.49\linewidth}{0.39\linewidth}{movs/A963_3D_main_clusters_v2.prc}
\includemovie[
  3Dc2c=-0.5560300946235657 -0.5524190664291382 -0.6210182309150696, 
  3Droo=6.000004177181491, 
  3Droll=27.22000797469873, 
  3Dbg=1. 1. 1., 
  3Dlights=CAD, 
  3Drender=Solid, 
	3Dviews2=movs/views_2192.tex,
       % 3Droo=7,3Dlights=CAD,
    poster=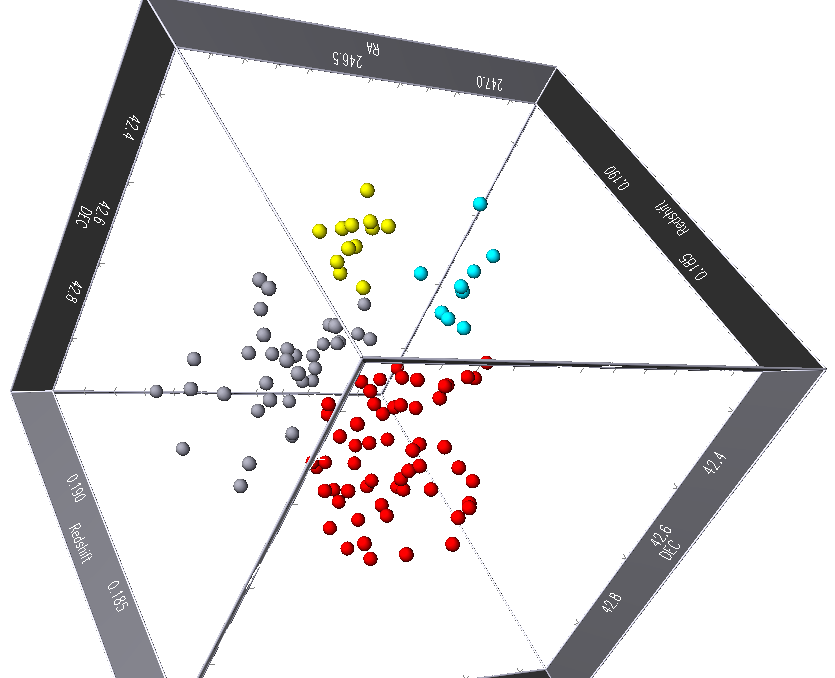,
    toolbar,
    label=pt2,
   % text={click to ativate},     
    ]{0.49\linewidth}{0.39\linewidth}{movs/A2192_3D_main_cluste_v2.prc}
\end{center} 
\caption{3D visualization of the distribution of galaxies in the main clusters. % ($0.16\leq z\leq0.22$). 
The galaxies are plotted in right ascension vs. declination vs. redshift. The left panel shows the massive cluster A963\_1 (red), its substructures A963\_1a and A963\_1b (coloured in turquoise and green respectivelly, see Sections~\ref{subsec:3DS}and~\ref{sec:env_sum}), and  A963\_1's smaller nearby companion, A963\_2 (blue). Smaller symbos represent the galaxies with less reliable redshifts \citep[including the cluster members from][]{LaveryHenry1994}.   %, and the cosmic sheet A963\_3 (green).
 The right-hand side shows A2192\_1 split into its substructures: A2192\_1a (red), A2192\_1b (gray), A2192\_1c (turquoise), and A2192\_1d (yellow). %and the groups A2192\_2 (green) and A2192\_3 (blue).
 %Note that  the redshift range in the case of A2192 has been slightly enlarged so that A2192\_3 is visible. In both panels, other field galaxies are plotted in grey symbols.  
This figure is three-dimensionally interactive in the online version of the paper, allowing the reader to change the magnification and viewing angle. The plots were constructed with the S2PLOT programming library \citep{Barnes2006}. A recent version of Adobe Acrobat Reader is required to display the content of this figure correctly. Click on each panel to activate the figure. Movies of the animated figures can also be found on the BUDHIES website  \url{http://www.astro.rug.nl/budhies/}, or in the online supplementary material. }
 \label{3d_plot}
\end{figure*}

%%%%%%AQUIIIII%%%%%

\section{Cluster substructure}
\label{sec:substructure}

Numerical dark matter simulations of galaxy groups and clusters in a $\Lambda$CDM universe predict that a significant fraction (30\%) of all systems should contain substructure \citep{KnebeMuller2000}. Because it reflects the dynamical state of a cluster, it is thus important to quantify the incidence of substructure, and to take it into account when studying galaxy properties as a function of environment. 
In \citet{Jaffe2012} we showed that A2192\_1 is a cluster with very clear spatially and dynamically distinct substructures inside it. Our analysis suggested that it is a cluster that is in the process of forming. 

There are many ways to detect substructure in clusters. A first approach is to study the Gaussianity of the velocity distribution. For example, from Figure~\ref{zhist} (middle-right panel) alone we can note a double-peak in A2192\_1 that already suggests the presence of substructure. However, velocity information alone does not necessarily reveal substructures within a cluster. Instead, it is important to look for deviations in the spatial and velocity distribution of galaxies simultaneously.

\subsection{A Three-dimensional view of the clusters}
\label{subsec:3dviews}

To study and identify the presence of substructure within the clusters, we looked for deviations in the spatial and/or velocity distribution of galaxies in each structure. One way to do this is to visually inspect 3-dimensional (3D)  maps of the galaxies' $\alpha$, $\delta$ and redshift  space.
Figure \ref{3d_plot} shows the 3D maps, centered on the richest structures in the survey. 

From this exercise, we can clearly see galaxy overdensities that are well separated in space and velocity. The most striking case is that of A2192\_1, because it shows clearly four separated substructures in space and velocity (coloured in red, gray, yellow and turquoise). This cluster has been thoroughly analyzed in \citet{Jaffe2012}, where we have also studied the distrubution of HI-detected galaxies around the substructures. % and summarized in Section~\ref{sec:env_sum}. 
A963\_1 on the other hand presents a different case. It is populated by more galaxies than A2192\_1 and their distribution in 3D space is more homogeneous. A clear feature is the ``finger'' of galaxies along the $z$-direction, that represents the cluster core, but in in the 3D distribution of this cluster there are no easily distinguishable substructures. In the following (Section~\ref{subsec:3DS}) we will complement the 3D view of the clusters with statistical substructure tests, and in Section~\ref{sec:env_sum} we summarize the results of our environment analysis.

\subsection{The Dressler-Shectman test}
\label{subsec:3DS}

To further test the presence of substructure in the clusters and groups, we carried out the Dressler-Shectman (DS) test  \citep{DresslerShectman1988}, that compares the \textit{local} velocity and velocity dispersion  for each galaxy with the \textit{global} values. To do this, we define ($\bar{v}_{\rm cl}$ and $\sigma_{\rm cl}$) as the mean velocity and velocity dispersion of the cluster/group, which is assumed to have $N_{\rm mem}$ galaxies. Then, for each galaxy \textit{i}, we select a subsample of galaxies containing the galaxy $i$, plus its nearest $N_{\rm nn}$ neighbors, and compute their mean velocity $\bar{v}^{i}_{\rm local}$ and velocity dispersion $\sigma^{i}_{\rm local}$. From these, we compute the individual galaxy deviations $\delta_{i}$, following:

\begin{equation}
\label{deltai}
\delta^{2}_{i}=\left(\frac{N_{\rm nn}+1}{\sigma^{2}_{\rm cl}}\right) \left[ (\bar{v}^{i}_{\rm local}  -  \bar{v}_{\rm cl})^{2} + (\sigma^{i}_{\rm local}  - \sigma_{\rm cl})^{2}  \right]
\end{equation}

We used $N_{\rm nn} = 10$ in our clusters and $N_{\rm nn} = \sqrt{N_{\rm mem}}$ in  groups with less than 20 members.

We performed two DS statistical tests: 

\begin{enumerate}
 \item The first is the ``critical value'' method, in which a $\Delta$-value is computed by:

\begin{equation}
\label{Delta}
\Delta=\sum_{i} \delta_{i}
\end{equation} 

\begin{figure*}
\begin{center}
  \includegraphics[width=0.49\textwidth]{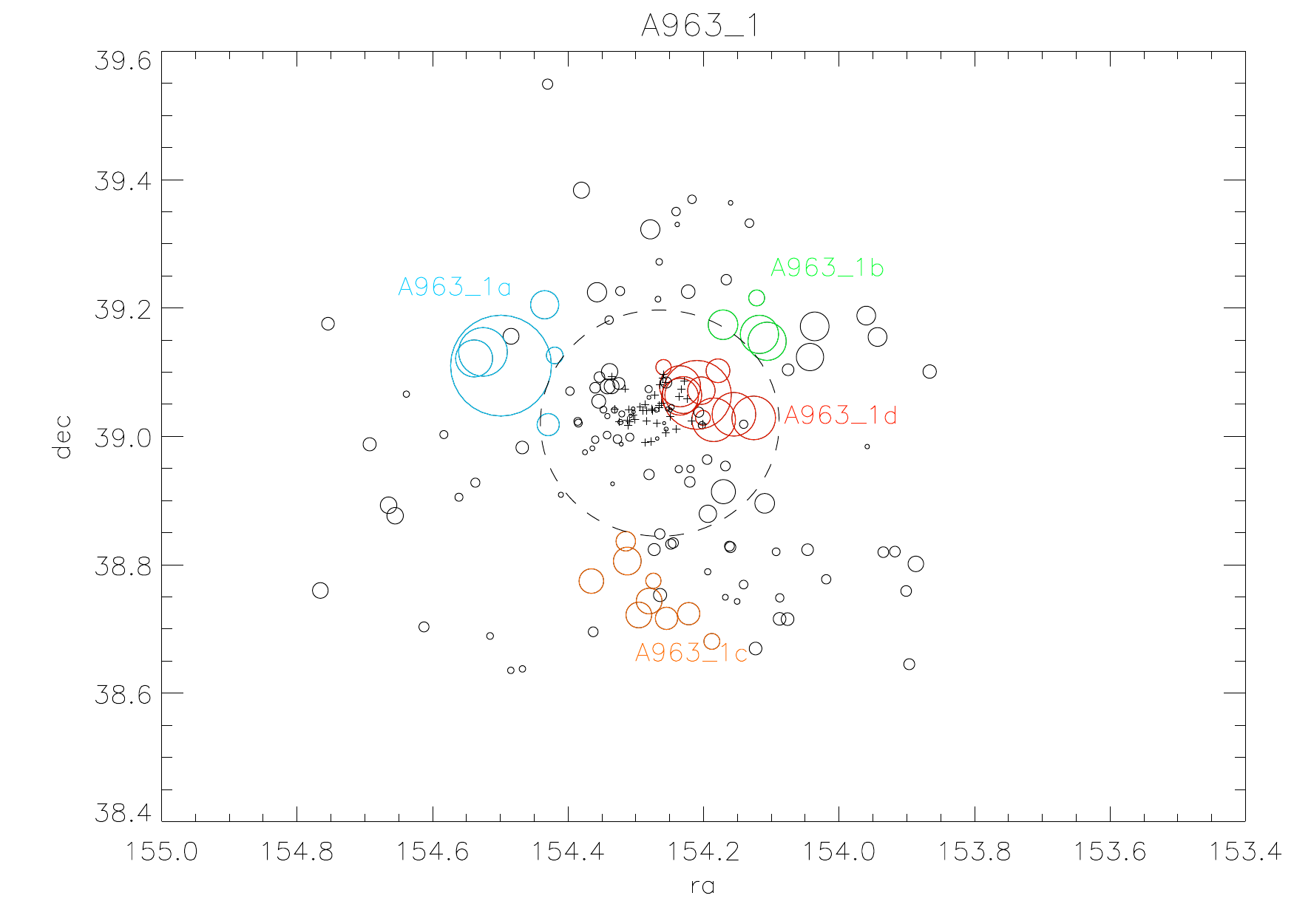}
  \includegraphics[width=0.49\textwidth]{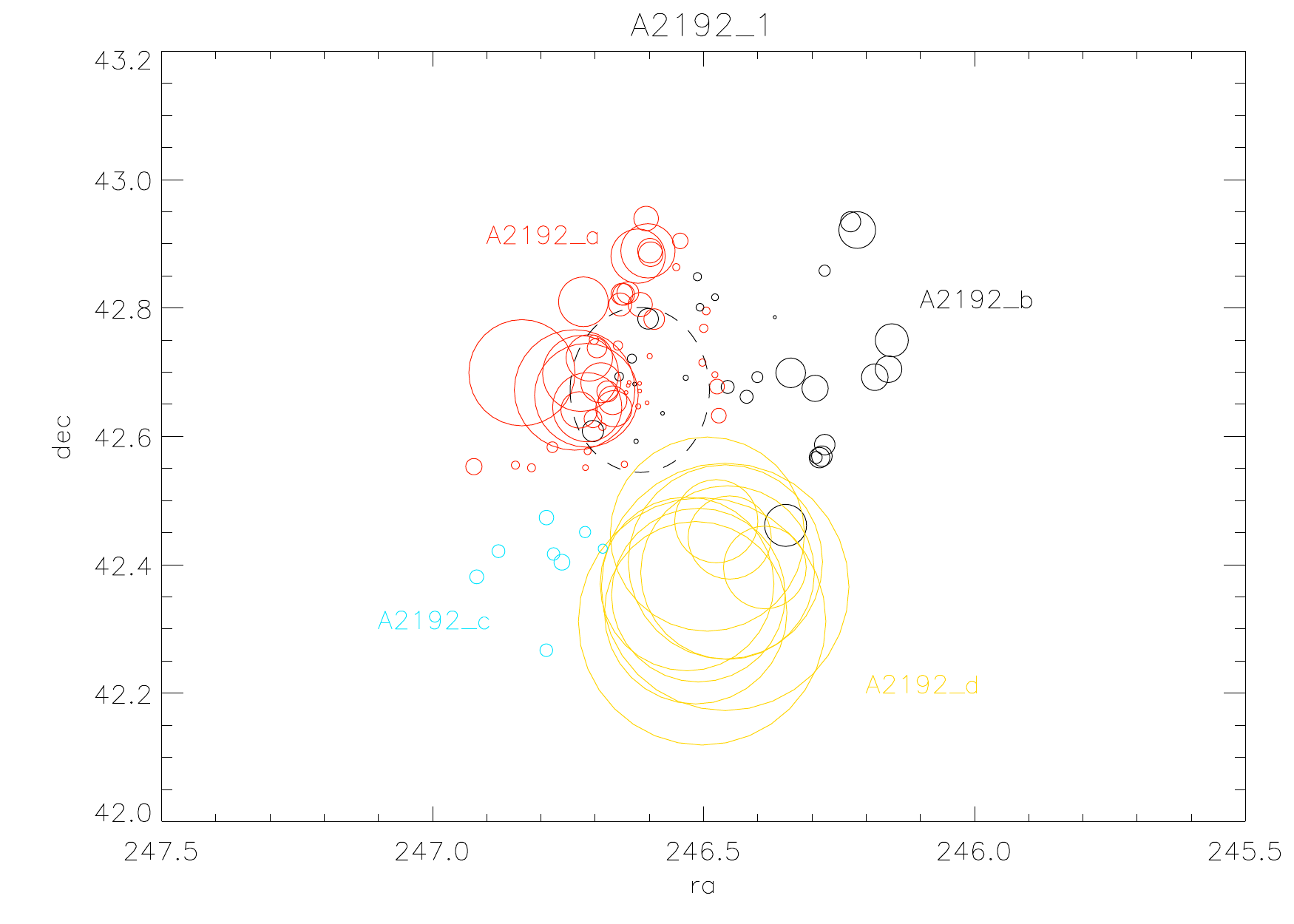}
\end{center} 
\caption{The  \citet{DresslerShectman1988} ``bubble-plot'' for the two main clusters in our sample. Galaxies are plotted with circles with a diameter that scales with  exp($\delta_{i}$). Colours indicate substructures identified. Dashed circles indicate the $R_{200}$ radius of the clusters. Only galaxies with good quality redshifts inside the CMD box were considered. The crosses in the the centre of A963 (left panel) correspond to additional cluster members from \citet[][and private communication]{LaveryHenry1994}. The ``bubble-plots'' for the other clusters/groups in our sample are shown in Appendix~\ref{ap:DS}.}
 \label{DS_bubble_plots}
\end{figure*}

\begin{table}
\begin{center}
 \caption{Results from the Dressler-Schectman test for substructure. Higher values of $\Delta_{\rm obs}/N_{\rm mem}$ indicate more likeliness to host substructure. Inversly, lower values of $P$ indicate a higher probability that the cluster/group has significant substructure. We adopt  $P\lesssim0.01$ to identify the cluster/groups with significant substructure.} 
\label{DS_table}
\begin{tabular}{cccc}
 \hline\\[-1mm]
Name of structure   & $N_{\rm mem}$ & $\Delta_{\rm obs}/N_{\rm mem}$ & $P$ \\
\hline\\[-2mm]
A963\_1		& 140	& 1.44	&	0.005	\\
A963\_2		& 21	& 1.65	&	0.27	\\
A963\_3		& 16	& 1.43	&	0.5	\\
\hline\\[-2mm]
A2192\_1	& 103	& 1.96	&	$<$0.001	\\
A2192\_2	& 15	& 1.91	&	0.015	\\
A2192\_3	& 11	& 1.79	&	0.011	\\
\hline\\
\end{tabular} 
\\ 
\end{center}
\end{table}

Then, a system is considered to have substructure if $\Delta$/$N_{\rm mem}>1$.
 \item The second method uses probabilities ($P$) rather than critical values. The $P$-values are computed by comparing the $\Delta$-value to ``shuffled'' $\Delta$-values, which are computed by randomly shuffling the observed velocities and reassigning these values to the member positions (i.e. Monte Carlo shuffling). The $P$-values are given by:

\begin{equation}
\label{Pvalue}
P=\sum(\Delta_{\rm shuffle} > \Delta_{\rm obs} / N_{\rm shuffle})
\end{equation} 
\end{enumerate}

Where $\Delta_{\rm shuffle}$ and $\Delta_{\rm obs}$ are computed following Equation \ref{Delta} and $N_{\rm shuffle}$ is the number of Monte Carlo shuffles performed, typically around 5000. A system is then considered to host substructure if it has a very small $P$-value ($\lesssim0.01$), as it is unlikely to obtain $\Delta_{\rm obs}$ randomly.

We applied both methods to the spectroscopic sample (see Table~\ref{DS_table}), but focus on the $P$ values to identify cluster/groups with substructure.  
We found that the main clusters,  A2192\_1 and A963\_1 have very small $P$ values ($\lesssim0.01$), which strongly suggests that they have significant substructure. 
The other clusters and groups have smaller, although non negligible, probabilities of hosting substructure (i.e. higher $P$ values).

Figures~\ref{DS_bubble_plots} and~\ref{DS_bubble_plots_cont}  further show the \citet{DresslerShectman1988} ``bubble-plots'' for each cluster/group in our sample. 
In these plots, each galaxy is marked by a circle with a size proportional to exp$(\delta_{i})$, so that galaxies with kinematics deviating significantly from the kinematics of the cluster can be easily identified. In our study, we considered significant deviations those with exp$(\delta_{i})\gtrsim 6$. 
In these figures the spectroscopic sample defined in Section~\ref{subsec:target} was utilized. Only high quality redshifts ($\geq-1$) were considered, which excludes literature data from \citet[][and private communication]{LaveryHenry1994}.

\begin{figure*}
\begin{center}
  \includegraphics[width=0.9\textwidth]{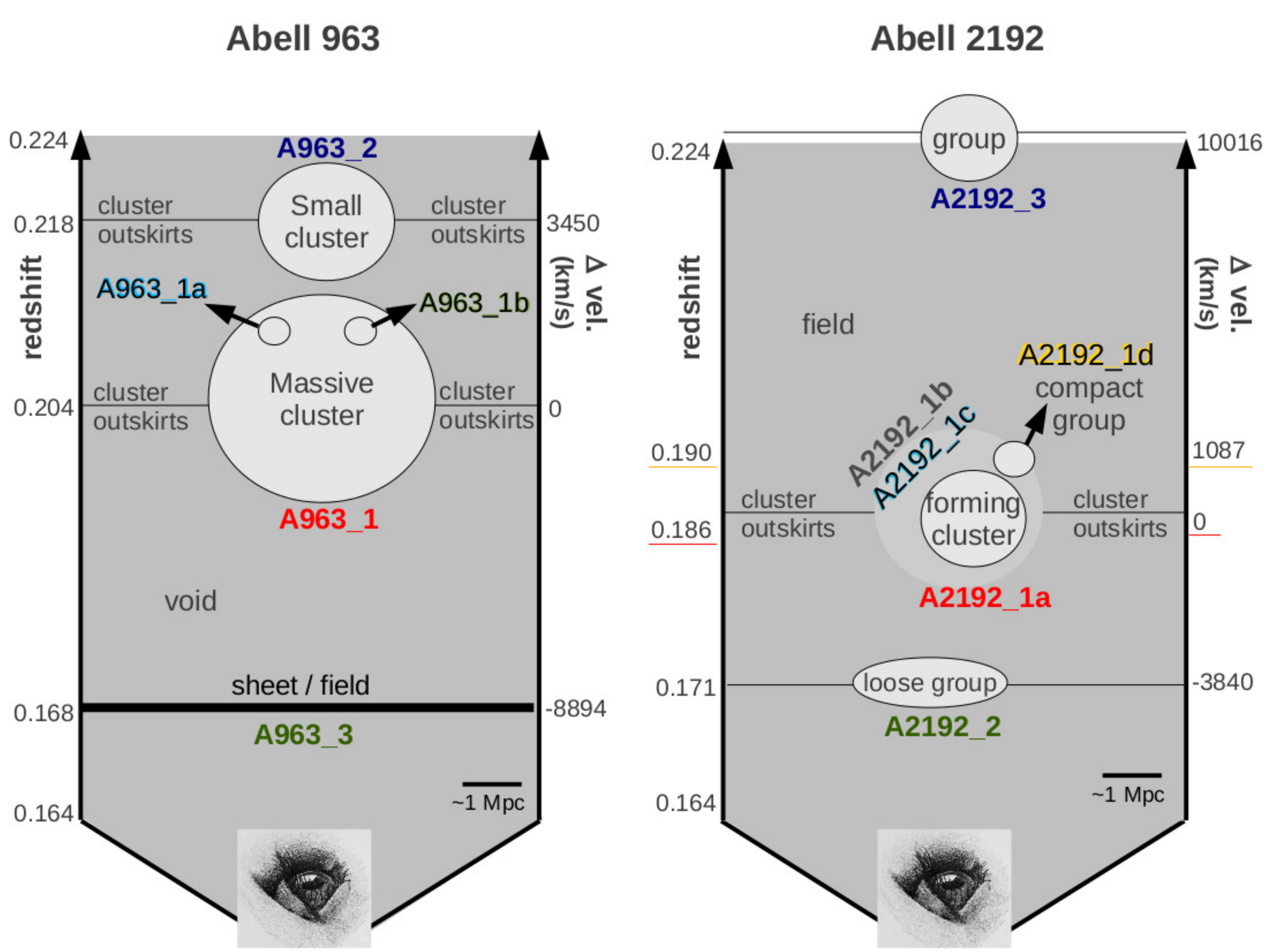} %schematic_environment_def_3}
\end{center} 
\caption{Schematic view of the different environments defined in this paper. The eyes looking upwards at the bottom of each panel indicate the line-of-sight. As a consequence redshift increases vertically (left axis). Only the redshift range of the HI observations is shown. Additionally, the axes on the right of each panel indicate the velocity difference ($\Delta \rm vel.$) between each structure and the main cluster (A963\_1 in the left panel and A2192\_1a in the right panel). The sizes of the clusters have been roughly scaled for comparison, in accordance with the 1 Mpc scale (at the main cluster's redshift) shown at the bottom-right (black solid line) of each panel.}
 \label{schema2}
\end{figure*}

%To identify possible individual substructures, 
We used the ``bubble-plots'' to confirm the substructures found in Figure~\ref{3d_plot} for A2192\_1  \citep[see also][]{Jaffe2012} and further identify less obvious substructures in A963\_1. First, we selected groups of galaxies with bigger circles in the bubble-plots (i.e. significant kinematic deviation from the cluster) in a similar manner as in \citet{BravoAlfaro2009}. In addition to this, once a substructure (i.e. a group of galaxies with similar-size circles) was identified, we examined the velocity distribution and discarded clear outliers. In this way we can be more certain that the substructures we identify are not only spatially concentrated, but also have consistent velocities.  For this reason, the substructures coloured in Figure~\ref{DS_bubble_plots} may have similar-sized uncoloured circles near them that are not considered part of the substructure (see for example the case of A963\_1b). This process confirmed the reality of the four substructures found in A2192\_1, and identified two poor groups with consistent velocities in the outskirts of A963\_1 (A963\_1a and A963\_1b, fully discussed in Section~\ref{sec:env_sum}).

Overall, the Dressler-Shectman test revealed that the substructure in A963\_1 is more sparsely distributed around the cluster outskirts (bigger -black and coloured- circles  outside $R_{200}$ in the left-hand side of Figure~\ref{DS_bubble_plots}), whilst for A2192\_1, the substructures found (coloured circles in the right-hand side of  Figure~\ref{DS_bubble_plots}) are very distinct, as we had found already in Figure~\ref{3d_plot} and \citet{Jaffe2012} from the 3D distribution of the galaxies.   

\section{Summary of the structures}
\label{sec:env_sum}

Our environmental analysis (c.f. Sections~\ref{subsec:3DS}  and~\ref{subsec:3dviews}) yielded a wide range of environments in the two surveyed volumes. 
Specifically,  we identified two clusters, one in each field, containing well-defined substructure within them. Together with these clusters, we found several foreground and background cluster/groups\footnote{As explained in Section\ref{subsec:redshifts}, we separate clusters from groups using a threshold cluster velocity dispersion value of 500km$s^{-1}$.} in the redshift range $0.164\leqslant z \leqslant0.224$, as illustrated in Figure~\ref{schema2}.  The main properties of the structures found are listed in Table~\ref{structures}. 

In the following, we combine all our results to  summarize the main characteristics of each structure.

\begin{figure*}
\begin{center}
  \includegraphics[width=0.49\textwidth]{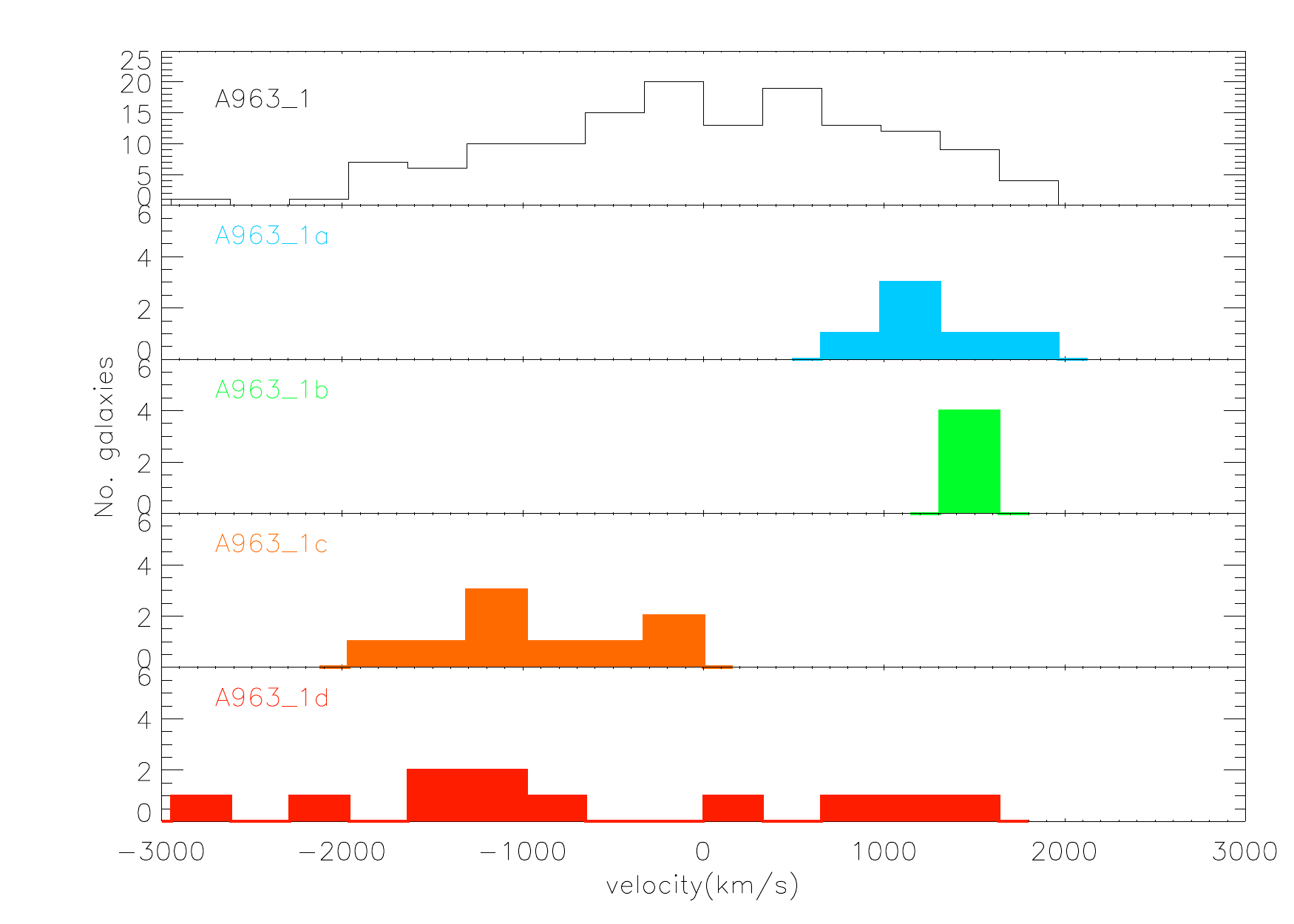}
  \includegraphics[width=0.49\textwidth]{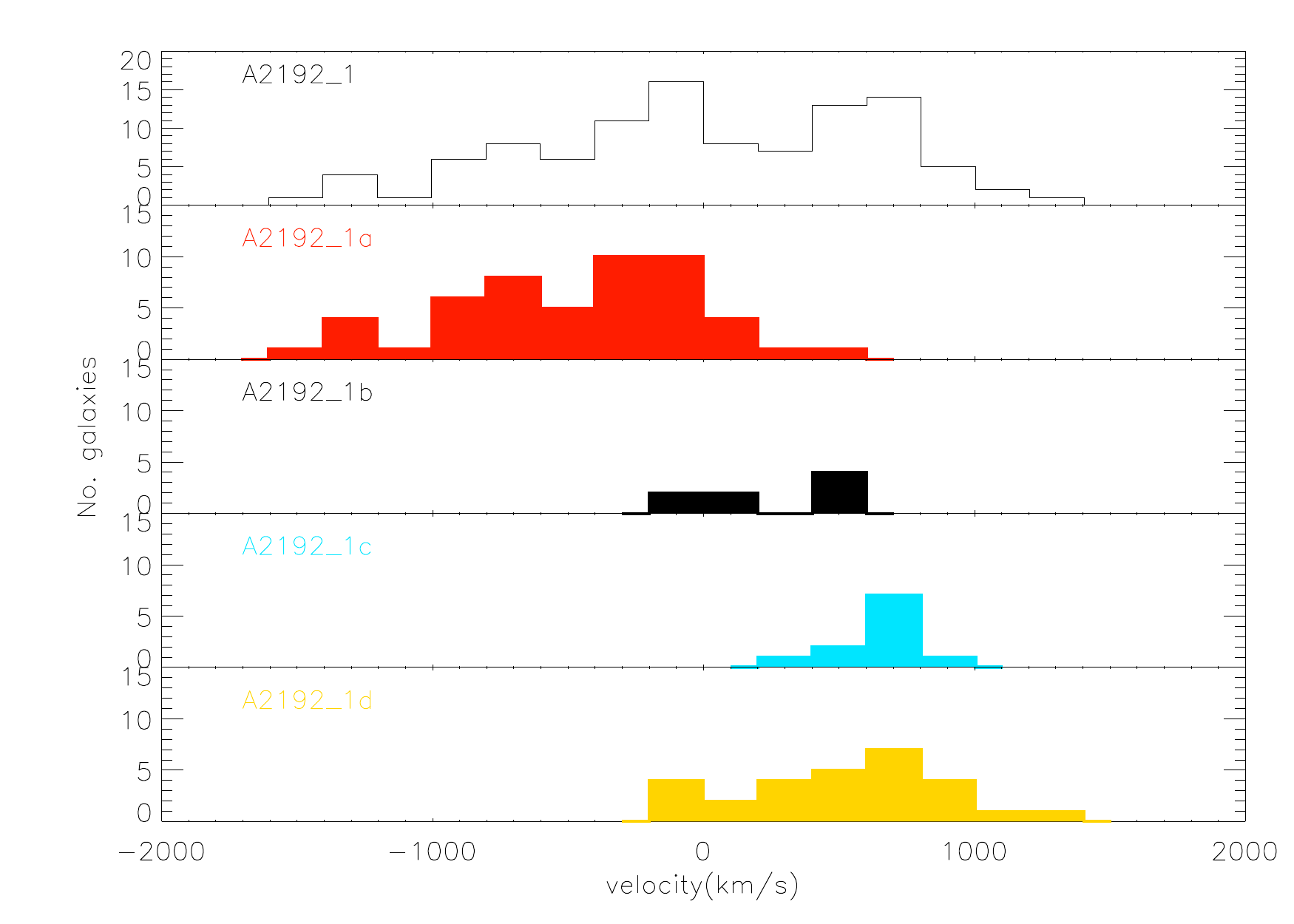}
\end{center} 
\caption{Velocity distribution of the substructures within A963\_1 and A2192\_1, identified in Figure~\ref{DS_bubble_plots}. For reference, the velocity distribution of all the galaxies in A963\_1 and A2192\_1 respectively are shown at the top. In the other panels,the colours and labels are as in Figure~\ref{DS_bubble_plots}. The spread in velocity of A963\_1c and A963\_1d suggests that they are not virialized galaxy groups, whilst the narrow velocity range of A963\_1a and A963\_1b strongly suggest the presence of a bound group of galaxies. In the case of the structures in A2192\_1, they all have a clearly separated velocity distributions.}
 \label{vel_s_a963_1}
\end{figure*}

\begin{itemize}
 \item \textbf{A963\_1:}  is the richest and most massive cluster in the surveyed volumes. Consisting of 141 spectroscopically confirmed galaxies (with $R<19.5$), it has a dynamical mass of $1.1\times10^{15} h^{-1}M_{\odot}$ and a virial radius of 1.55 h$^{-1}$Mpc (both estimated from the measured $\sigma_{\rm cl}=993$km s$^{-1}$). 
A963\_1 has previously been studied both in X-rays and using weak lensing \citep[][]{Allen2003,Smith2005}.
The cluster has been classified as ``relaxed'', as it shows very regular and centrally concentrated  X-ray  and mass morphologies (on a central cD galaxy) \citep[see Fig. 6 of ][]{Smith2005}, which suggests a low level of substructure. Its total mass has been estimated to be 3.5$\pm0.3\times10^{14}h^{-1}M_{\odot}$ from the lensing analysis, which is lower than our $\sigma_{\rm cl}$-derived measure. Moreover, its total X-ray luminosity is $L_{X}=3.4\pm1\times10^{44}h^{-2}$erg/s, and the X-ray estimated $R_{200}$ is 1.2$\pm0.1 h^{-1}$Mpc, which is also smaller than the dynamical value we measure ($=1.55 h^{-1}$Mpc, see Table~\ref{structures}). Our analysis, combined with the X-ray observations, suggests that the ``relaxed'' part of the cluster is at its core, while the substructure dominates the outskirts. 

The DS test in A963\_1 yielded $\Delta/N_{\rm mem}=1.44$ and $P=0.005$ (c.f. Table~\ref{DS_table}), which strongly indicates it has substructure. Figure~\ref{DS_bubble_plots} reveals a spread population of galaxies surrounding the cluster with evident velocity and spatial offsets from the cluster (see bigger -black and coloured- circles surrounding the cluster centre), suggesting it might be a population of infalling field galaxies. Moreover, there are 4 substructures (coloured and labeled in the figure) that could represent accreted galaxy groups in the process of destruction. This seems plausible given that they lie within the turnaround radius (4.57$\times R_{200}$).  To verify the reality of these groups, we inspected their velocity distributions, as shown in Figure~\ref{vel_s_a963_1}. We find that only 2 groups, namely A963\_1a and A963\_1b have tight velocity distributions (at 1230 and 1548 km s$^{-1}$ from the main cluster respectively), indicating they are likely bound groups. However, in A963\_1c and A963\_1d  there is not a clear separation in the velocity space alone. In fact, their broad velocity distributions indicate that they are not gravitationally bond groups.  
We thus conclude that A963\_1 is a cluster with significant substructure, including  two poor groups at $\gtrsim R_{200}$, but dominated by a spread population of infalling galaxies in the cluster's outskirts.

\item \textbf{A963\_2:} is a $1.4\times10^{14} h^{-1}M_{\odot}$ ($\sigma_{\rm cl}=507$km s$^{-1}$) cluster located at the same position in the sky as A963\_1, but  at slightly higher redshift. 
Although it is tempting to think that A963\_2 is in the process of falling into A963\_1, it is important to note that A963\_2 lies outside the turnaround radius of the main cluster. In fact, 
A963\_2 has a velocity difference of $\sim3450$km s$^{-1}$ from A963\_1's centre  (over 3 times the $\sigma_{\rm cl}$ of A963\_1), so it is unlikely that these clusters are interacting.   

The velocity distribution of this cluster resembles a Gaussian, although it also shows  some hints of asymmetry. In particular, it shows a prominent and narrow velocity peak at the central velocity with a normal distribution underneath it. If this feature is real, it could be interpreted as a cosmic  sheet at the peak's redshift that contains an intermediate-size cluster.  
Moreover, the substructure analysis in A963\_2 does not give strong evidence of substructure.

\item \textbf{A963\_3:} This family of galaxies with a narrow velocity distribution but very spread out in space,  is likely to be a cosmic sheet and not a galaxy group. There is a very low probability that A963\_3 contains substructure, supporting the idea that this group is just a population of field galaxies at the same redshift.

 \item \textbf{A2192\_1:} is a forming cluster with significant substructure. This cluster has been fully explored in \citet{Jaffe2012}. In short, A2192\_1 is composed by a $1.6\times10^{14} h^{-1}M_{\odot}$ ($\sigma_{\rm cl}=530$km s$^{-1}$) cluster, A2192\_1a, which is the forming cluster itself that coincides with (weak) X-ray emission. 
Surrounding the cluster, we find a compact group, A2192\_1d, and a scattered population of  ``field-like'' galaxies A2192\_1b and A2192\_1c. As explained in \citet{Jaffe2012}, A2192\_1b and A2192\_1c  are ``field-like'' because they are spread in space and their galaxy population resembles that of the field. These substructures, although unlikely to be bound, are considered part of A2192\_1 since they are well within the turnaround radius. 
Supporting our findings, the Dressler-Shectman ``bubble-plot'' (right-hand side of Figure~\ref{DS_bubble_plots}) revealed the same substructures found  in Figure~\ref{3d_plot} \citep[see also  Figure~2  of ][]{Jaffe2012}.

In contrast with A963\_1, A2192\_1 has very weak X-ray emission. However, the existent emission coincides with a group of early-type galaxies at the core of A2192\_1a. Moreover,  the X-ray luminosity \citep[$L_{X}\simeq7\times10^{43}h^{-2}$ erg/s; ][]{Voges99} is consistent with the derived dynamical mass.

 \item \textbf{A2192\_2:} with a dynamical mass of $9.9\times10^{13} h^{-1}M_{\odot}$ ($\sigma_{\rm cl}=447$km s$^{-1}$), this galaxy group is clearly separated in velocity space from the others in the field. Although its mass is very similar to that of the low-mass cluster A963\_2, A2192\_2  has a $\sigma_{\rm cl}$ below our threshold limit, so we have classified it as a group. Moreover it is considerably spread in space and has substructure, which makes it a loose group. 
 
 \item \textbf{A2192\_3:} is a $4.21\times10^{13} h^{-1}M_{\odot}$ ($\sigma_{\rm cl}=340$km s$^{-1}$) group of galaxies with little substructure. 

Table~\ref{structures} lists the main characteristics of the structures  described above. Other cluster/groups outside the surveyed volume, but identified by our spectroscopic campaign, are listed in Table~\ref{structures_2} for completeness.

\end{itemize}

\begin{table*}
\begin{center}
 \caption{Clusters, groups and other structures identified in the fields of A963 and A2192 inside the redshift range of the HI-observations. The columns indicate the field, structure, central redshift ($z_{c}$), number of members in the reduced sample (and -for the main clusters- number of members within $R_{200}$), cluster velocity dispersion, $R_{200}$, and comments.} 
\label{structures}
\begin{tabular}{llccccl}
\hline\\[-1mm]
Field 	& Name of structure   &  $z_{c}$   & No. members	& $\sigma_{\rm cl}$ & R$_{200}$ & Remarks\\ 
 	& 		      &                      & ($R<19.4$mag)	& (km s$^{-1}$) 	    & (h$^{-1}$Mpc)	 & \\ 
\hline\\[-2mm]
A963	& \textbf{A963\_1}		& 0.2039 		& 141 (67)	& 993$\pm56$	&1.55	& Main cluster in A963	\\
	&  ~~A963\_1a 			& 0.2089		& 7 			& --				& --	 & Small substructure in A963\_1	\\
	&  ~~A963\_1b 			& 0.2103		& 4 			& --				& --	 & Small substructure in A963\_1	\\	
	& \textbf{A963\_2}			& 0.2178		& 21 		& 507$\pm158$	&0.79	& Intermediate-size cluster near A963\_1	\\
	& \textbf{A963\_3}			& 0.1682		& 15 		& --				&--	& Cosmic sheet, very narrow in z and spread in $\alpha - \delta$	\\
	&\textbf{A963 field}		& $0.164\leqslant z \leqslant0.224$		& 30		& --		&--	& Galaxies not belonging to any cluster/group	\\
\hline\\[-2mm]
A2192	&  \textbf{A2192\_1}	& 0.1876 		& 103 (39)	& 645$\pm42$ 	& 1.02	 & Main cluster in A2192 \citep[see][]{Jaffe2012}	\\	
	&  ~~A2192\_1a 			& 0.1859		& 50 		& 530$\pm56$	& 0.84	 & Substructure of A2192\_1	\\	
	&  ~~A2192\_1b			& 0.1898		& 29 		& --				& --		 & ``Field-like'' substructure of A2192\_1		\\
	&  ~~A2192\_1c			& 0.1881		& 8 			& --				& --		 & ``Field-like'' substructure of A2192\_1		\\	
	&  ~~A2192\_1d			& 0.1902		& 11 		& 161$\pm52$	& 0.25	 & Compact group, substructure of A2192\_1		\\
	&  \textbf{A2192\_2}		& 0.1707		& 17 		& 477$\pm103$	&  0.71  	& Loose group 	\\
	&  \textbf{A2192\_3}*		& 0.2255		& 30 		& 340$\pm71$	& 0.53 	 & Galaxy group 	\\	
	&\textbf{A2192 field}		& $0.164\leqslant z\leqslant0.224$		& 20		& --	&--	 &Galaxies not belonging to any cluster/group	\\
\hline\\
\end{tabular}
$^{*}$ This group spans a redshift range that exceeds the limit of the HI survey, and hence suffers from incompleteness.
\end{center}

\end{table*}

\section{Conclusions}
\label{sec:concl}

%We have carried out a complete HI analysis of a Blind Ultra Deep HI Environmental Survey (BUDHIES) of two fields at $0.164\leq z \leq0.224$  (Deshev et~al. in preparation). 
With the WSRT, BUDHIES has detected HI in over 150 galaxies in and around 2 Abell clusters at z~0.2  (Deshev et al. in preparation). 
%
%The fields contain two Abell clusters: 963 and 2192. 
As part of BUDHIES, we have also carried out spectroscopic observations of the galaxies in the two surveyed volumes with AF2+WYFFOS on the WHT. 
%we have carried out spectroscopic observations with AF2+WYFFOS/WHT of galaxies in the two surveyed volumes. 
In this paper we provide details of the spectroscopic target selection, observations, and data reduction. 
The observations allowed us to measure many redshifts that mapped the cosmic large-scale structure in these volumes and thus characterize the environment of the HI-detected galaxies in the surveyed volume. 
We present data tables containing positions, redshifts, equivalent widths of the [OII] emission, and $B-$ and $R-$band magnitudes for galaxies in the volumes. 

The redshift distribution revealed several structures in each of the surveyed volume, ranging from massive clusters to small groups, cosmic sheets and voids. 
We further present an environment classification scheme in which we consider ``cluster'' galaxies those within $R_{200}$, ``outskirts'' the region between $R_{200}$ and the turnaround radius, and ``field'' all the galaxies beyond the turnaround radius.

By performing detailed substructure tests, we find that both main clusters (A963\_1 and A2192\_1) show a high degree of substructure within their turnaround radii, suggesting they are actively accreting galaxies from smaller groups or the field population. 
In particular, A963\_1 shows weaker evidence for group accretion, and stronger evidence for accretion of (originally) field galaxies that are well spread in space and velocity. On the contrary, A2192\_1 has very distinct substructures, strongly implying that the cluster is in the process of forming from the accretion of galaxy groups.  
The other cluster/groups in the survey have less complex substructure.  

The diversity of environments found in the surveyed volumes, together with the unprecedented multi-wavelength dataset, makes BUDHIES an ideal laboratory to study environmental dependent processes such as strangulation, tidal and ram pressure stripping and the importance of group pre-processing in galaxy evolution. 
%%%%%%%%%%%%%%%%%%%%%%%%%%%%%%%%%%%%%%%%%%%%%%%%%%%%%%%%%%%%%%%%

\section*{Acknowledgments}

YLJ and BMP acknowledge financial support from ASI through contract I/099/10/0, and FONDECYT grant No. 3130476. 
This work was supported in part by the National Science Foundation under grant No. 1009476 to Columbia University. 
We are grateful for support from a Da Vinci Professorship at the Kapteyn Institute. 
We thank Richard Jackson and Ian Skillen for their support on the new AF2 pipeline, and Daniela Bettoni for useful discussions. YLJ is grateful to Graeme Candlish for asistance creating the 3D plots.

%\begin{thebibliography}{}

%\bibliography{/home/yara/Desktop/refs}{}

\begin{thebibliography}{}

\bibitem[\protect\citeauthoryear{{Abramson}, {Kenney}, {Crowl}, {Chung}, {van
  Gorkom}, {Vollmer} \& {Schiminovich}}{{Abramson} et~al.}{2011}]{Abramson2011}
{Abramson} A.,  {Kenney} J.~D.~P.,  {Crowl} H.~H.,  {Chung} A.,  {van Gorkom}
  J.~H.,  {Vollmer} B.,    {Schiminovich} D.,  2011, \aj, 141, 164

\bibitem[\protect\citeauthoryear{{Allen}, {Schmidt}, {Fabian} \&
  {Ebeling}}{{Allen} et~al.}{2003}]{Allen2003}
{Allen} S.~W.,  {Schmidt} R.~W.,  {Fabian} A.~C.,    {Ebeling} H.,  2003,
  \mnras, 342, 287

\bibitem[\protect\citeauthoryear{{Barnes}, {Fluke}, {Bourke} \&
  {Parry}}{{Barnes} et~al.}{2006}]{Barnes2006}
{Barnes} D.~G.,  {Fluke} C.~J.,  {Bourke} P.~D.,    {Parry} O.~T.,  2006,
  \pasa, 23, 82

\bibitem[\protect\citeauthoryear{{Beers}, {Flynn} \& {Gebhardt}}{{Beers}
  et~al.}{1990}]{bfg90}
{Beers} T.~C.,  {Flynn} K.,    {Gebhardt} K.,  1990, \aj, 100, 32

\bibitem[\protect\citeauthoryear{{Bell}}{{Bell}}{2007}]{Bell2007}
{Bell} M.~B.,  2007, \apjl, 667, L129

\bibitem[\protect\citeauthoryear{{Bolzonella},et~al.,}{{Bolzonella} et~al.}{2010}]{Bolzonella2010}
{Bolzonella} M., et~al., 2010,
  \aap, 524, A76+


\bibitem[\protect\citeauthoryear{{Bravo-Alfaro}, {Caretta}, {Lobo}, {Durret} \&
  {Scott}}{{Bravo-Alfaro} et~al.}{2009}]{BravoAlfaro2009}
{Bravo-Alfaro} H.,  {Caretta} C.~A.,  {Lobo} C.,  {Durret} F.,    {Scott} T.,
  2009, \aap, 495, 379

\bibitem[\protect\citeauthoryear{{Bravo-Alfaro}, {Cayatte}, {van Gorkom} \&
  {Balkowski}}{{Bravo-Alfaro} et~al.}{2000}]{BravoAlfaro2000}
{Bravo-Alfaro} H.,  {Cayatte} V.,  {van Gorkom} J.~H.,    {Balkowski} C.,
  2000, \aj, 119, 580

\bibitem[\protect\citeauthoryear{{Bravo-Alfaro}, {Cayatte}, {van Gorkom} \&
  {Balkowski}}{{Bravo-Alfaro} et~al.}{2001}]{BravoAlfaro2001}
{Bravo-Alfaro} H.,  {Cayatte} V.,  {van Gorkom} J.~H.,    {Balkowski} C.,
  2001, \aap, 379, 347

\bibitem[\protect\citeauthoryear{{Butcher} \& {Oemler}}{{Butcher} \&
  {Oemler}}{1978}]{bo78}
{Butcher} H.,  {Oemler} A.,  1978, \apj, 219, 18

\bibitem[\protect\citeauthoryear{{Butcher}, {Wells} \& {Oemler} Jr.}{{Butcher}
  et~al.}{1983}]{Butcher1983}
{Butcher} H.,  {Wells} D.~C.,    {Oemler} Jr. A.,  1983, \apjs, 52, 183

\bibitem[\protect\citeauthoryear{{Carlberg}, {Yee}, {Ellingson}, {Abraham},
  {Gravel}, {Morris} \& {Pritchet}}{{Carlberg} et~al.}{1996}]{Carlberg1996}
{Carlberg} R.~G.,  {Yee} H.~K.~C.,  {Ellingson} E.,  {Abraham} R.,  {Gravel}
  P.,  {Morris} S.,    {Pritchet} C.~J.,  1996, \apj, 462, 32

\bibitem[\protect\citeauthoryear{{Cayatte}, {van Gorkom}, {Balkowski} \&
  {Kotanyi}}{{Cayatte} et~al.}{1990}]{Cayatte1990}
{Cayatte} V.,  {van Gorkom} J.~H.,  {Balkowski} C.,    {Kotanyi} C.,  1990,
  \aj, 100, 604

\bibitem[\protect\citeauthoryear{{Chung}, {van Gorkom}, {Kenney}, {Crowl} \&
  {Vollmer}}{{Chung} et~al.}{2009}]{Chung2009}
{Chung} A.,  {van Gorkom} J.~H.,  {Kenney} J.~D.~P.,  {Crowl} H.,    {Vollmer}
  B.,  2009, \aj, 138, 1741

\bibitem[\protect\citeauthoryear{{Chung}, {van Gorkom}, {Kenney} \&
  {Vollmer}}{{Chung} et~al.}{2007}]{Chung2007}
{Chung} A.,  {van Gorkom} J.~H.,  {Kenney} J.~D.~P.,    {Vollmer} B.,  2007,
  \apjl, 659, L115

\bibitem[\protect\citeauthoryear{{Crowl}, {Kenney}, {van Gorkom} \&
  {Vollmer}}{{Crowl} et~al.}{2005}]{crowl05a}
{Crowl} H.~H.,  {Kenney} J.~D.~P.,  {van Gorkom} J.~H.,    {Vollmer} B.,  2005,
  \aj, 130, 65

\bibitem[\protect\citeauthoryear{{De Lucia}, {Weinmann}, {Poggianti},
  {Arag{\'o}n-Salamanca} \& {Zaritsky}}{{De Lucia} et~al.}{2012}]{delucia2012}
{De Lucia} G.,  {Weinmann} S.,  {Poggianti} B.~M.,  {Arag{\'o}n-Salamanca} A.,
    {Zaritsky} D.,  2012, \mnras, 423, 1277

\bibitem[\protect\citeauthoryear{{Desai}, {Dalcanton}, {Arag{\'o}n-Salamanca},
  {Jablonka}, {Poggianti}, {Gogarten}, {Simard}, {Milvang-Jensen}, {Rudnick},
  {Zaritsky}, {Clowe}, {Halliday}, {Pell{\'o}}, {Saglia} \& {White}}{{Desai}
  et~al.}{2007}]{Desai2007}
{Desai} V.,  {Dalcanton} J.~J.,  {Arag{\'o}n-Salamanca} A.,  {Jablonka} P.,
  {Poggianti} B.,  {Gogarten} S.~M.,  {Simard} L.,  {Milvang-Jensen} B.,
  {Rudnick} G.,  {Zaritsky} D.,  {Clowe} D.,  {Halliday} C.,  {Pell{\'o}} R.,
  {Saglia} R.,    {White} S.,  2007, \apj, 660, 1151

\bibitem[\protect\citeauthoryear{{Dressler}}{{Dressler}}{1980}]{Dressler1980}
{Dressler} A.,  1980, \apj, 236, 351

\bibitem[\protect\citeauthoryear{{Dressler}, {Oemler} Jr., {Couch}, {Smail},
  {Ellis}, {Barger}, {Butcher}, {Poggianti} \& {Sharples}}{{Dressler}
  et~al.}{1997}]{Dressler1997}
{Dressler} A.,  {Oemler} Jr. A.,  {Couch} W.~J.,  {Smail} I.,  {Ellis} R.~S.,
  {Barger} A.,  {Butcher} H.,  {Poggianti} B.~M.,    {Sharples} R.~M.,  1997,
  \apj, 490, 577

\bibitem[\protect\citeauthoryear{{Dressler} \& {Shectman}}{{Dressler} \&
  {Shectman}}{1988}]{DresslerShectman1988}
{Dressler} A.,  {Shectman} S.~A.,  1988, \aj, 95, 985

\bibitem[\protect\citeauthoryear{{Ellingson}, {Lin}, {Yee} \&
  {Carlberg}}{{Ellingson} et~al.}{2001}]{Ellingson2001}
{Ellingson} E.,  {Lin} H.,  {Yee} H.~K.~C.,    {Carlberg} R.~G.,  2001, \apj,
  547, 609

\bibitem[\protect\citeauthoryear{{Fadda}, {Biviano}, {Marleau},
  {Storrie-Lombardi} \& {Durret}}{{Fadda} et~al.}{2008}]{Fadda2008}
{Fadda} D.,  {Biviano} A.,  {Marleau} F.~R.,  {Storrie-Lombardi} L.~J.,
  {Durret} F.,  2008, \apjl, 672, L9

\bibitem[\protect\citeauthoryear{{Fasano}, {Poggianti}, {Couch}, {Bettoni},
  {Kj{\ae}rgaard} \& {Moles}}{{Fasano} et~al.}{2000}]{fasano2000}
{Fasano} G.,  {Poggianti} B.~M.,  {Couch} W.~J.,  {Bettoni} D.,
  {Kj{\ae}rgaard} P.,    {Moles} M.,  2000, \apj, 542, 673

\bibitem[\protect\citeauthoryear{{Finn}, {Zaritsky}, {McCarthy}, {Poggianti},
  {Rudnick}, {Halliday}, {Milvang-Jensen}, {Pell{\'o}} \& {Simard}}{{Finn}
  et~al.}{2005}]{Finn2005}
{Finn} R.~A.,  {Zaritsky} D.,  {McCarthy} D.~W.,  {Poggianti} B.,  {Rudnick}
  G.,  {Halliday} C.,  {Milvang-Jensen} B.,  {Pell{\'o}} R.,    {Simard} L.,
  2005, \apj, 630, 206

\bibitem[\protect\citeauthoryear{{Freeland}, {Stilp} \& {Wilcots}}{{Freeland}
  et~al.}{2009}]{Freeland2009}
{Freeland} E.,  {Stilp} A.,    {Wilcots} E.,  2009, \aj, 138, 295

\bibitem[\protect\citeauthoryear{{Giovanelli}, et~al.,}{{Giovanelli} et~al.} {2005}] {Giovanelli2005}
{Giovanelli} R.,  et~al.,  2005, \aj, 130, 2598


\bibitem[\protect\citeauthoryear{{G{\'o}mez} et~al.,}{{G{\'o}mez}
  et~al.}{2003}]{gomez03}
{G{\'o}mez} P.~L.,  et~al., 2003, \apj, 584, 210

\bibitem[\protect\citeauthoryear{{Haines}, {Gargiulo}, {La Barbera},
  {Mercurio}, {Merluzzi} \& {Busarello}}{{Haines} et~al.}{2007}]{Haines2007}
{Haines} C.~P.,  {Gargiulo} A.,  {La Barbera} F.,  {Mercurio} A.,  {Merluzzi}
  P.,    {Busarello} G.,  2007, \mnras, 381, 7

\bibitem[\protect\citeauthoryear{{Harrison}}{{Harrison}}{1974}]{Harrison1974}
{Harrison} E.~R.,  1974, \apjl, 191, L51

\bibitem[\protect\citeauthoryear{{Hibbard}, {van der Hulst}, {Barnes} \&
  {Rich}}{{Hibbard} et~al.}{2001}]{Hibbard2001}
{Hibbard} J.~E.,  {van der Hulst} J.~M.,  {Barnes} J.~E.,    {Rich} R.~M.,
  2001, \aj, 122, 2969


\bibitem[\protect\citeauthoryear{{Jaff{\'e}}, {Arag{\'o}n-Salamanca}, {Kuntschner}, {Bamford},  {Hoyos},  {De Lucia}, {Halliday}, {Milvang-Jensen}, {Poggianti},  {Rudnick},  {Saglia}, {Sanchez-Blazquez} \& {Zaritsky}}{{Jaff{\'e}}  et~al.}{2011}]{Jaffe2011b}{Jaff{\'e}} Y.~L., {Arag{\'o}n-Salamanca} A., {Kuntschner} H., {Bamford} S.,  {Hoyos} C., {De Lucia} G., {Halliday} C., {Milvang-Jensen} B., {Poggianti}
  B., {Rudnick} G., {Saglia} R.~P., {Sanchez-Blazquez} P., {Zaritsky} D.,  2011, \mnras, 417, 1996

\bibitem[\protect\citeauthoryear{{Jaff{\'e}}, {Poggianti}, {Verheijen},
  {Deshev} \& {van Gorkom}}{{Jaff{\'e}} et~al.}{2012}]{Jaffe2012}
{Jaff{\'e}} Y.~L.,  {Poggianti} B.~M.,  {Verheijen} M.~A.~W.,  {Deshev} B.~Z.,
    {van Gorkom} J.~H.,  2012, \apjl, 756, L28

\bibitem[\protect\citeauthoryear{{Kapferer}, {Sluka}, {Schindler}, {Ferrari} \&
  {Ziegler}}{{Kapferer} et~al.}{2009}]{Kapferer2009}
{Kapferer} W.,  {Sluka} C.,  {Schindler} S.,  {Ferrari} C.,    {Ziegler} B.,
  2009, \aap, 499, 87

\bibitem[\protect\citeauthoryear{{Kauffmann}, {Colberg}, {Diaferio} \&
  {White}}{{Kauffmann} et~al.}{1999}]{Kauffmann1999}
{Kauffmann} G.,  {Colberg} J.~M.,  {Diaferio} A.,    {White} S.~D.~M.,  1999,
  \mnras, 307, 529

\bibitem[\protect\citeauthoryear{{Kenney}, {van Gorkom} \& {Vollmer}}{{Kenney}
  et~al.}{2004}]{Kenney2004}
{Kenney} J.~D.~P.,  {van Gorkom} J.~H.,    {Vollmer} B.,  2004, \aj, 127, 3361

\bibitem[\protect\citeauthoryear{{Kern}, {Kilborn}, {Forbes} \&
  {Koribalski}}{{Kern} et~al.}{2008}]{Kern2008}
{Kern} K.~M.,  {Kilborn} V.~A.,  {Forbes} D.~A.,    {Koribalski} B.,  2008,
  \mnras, 384, 305

\bibitem[\protect\citeauthoryear{{Knebe} \& {M{\"u}ller}}{{Knebe} \&
  {M{\"u}ller}}{2000}]{KnebeMuller2000}
{Knebe} A.,  {M{\"u}ller} V.,  2000, \aap, 354, 761

\bibitem[\protect\citeauthoryear{{Kodama} \& {Bower}}{{Kodama} \&
  {Bower}}{2001}]{KodamaBower2001}
{Kodama} T.,  {Bower} R.~G.,  2001, \mnras, 321, 18

\bibitem[\protect\citeauthoryear{{Lavery} \& {Henry}}{{Lavery} \&
  {Henry}}{1994}]{LaveryHenry1994}
{Lavery} R.~J.,  {Henry} J.~P.,  1994, \apj, 426, 524

\bibitem[\protect\citeauthoryear{{Lewis} et~al.,}{{Lewis}
  et~al.}{2002}]{lewis02}
{Lewis} I.,  et~al., 2002, \mnras, 334, 673

\bibitem[\protect\citeauthoryear{{Mahajan}, {Raychaudhury} \&
  {Pimbblet}}{{Mahajan} et~al.}{2012}]{Mahajan2012}
{Mahajan} S.,  {Raychaudhury} S.,    {Pimbblet} K.~A.,  2012, \mnras, 427, 1252

\bibitem[\protect\citeauthoryear{{Meyer} et~al.,}{{Meyer} et~al.} {2004}]{Meyer2004}
{Meyer} M.~J.,  et~al.,  2004, \mnras, 350, 1195

\bibitem[\protect\citeauthoryear{{Mulchaey}}{{Mulchaey}}{2000}]{Mulchaey2000}
{Mulchaey} J.~S.,  2000, \araa, 38, 289

\bibitem[\protect\citeauthoryear{{Oesch} et~al.,}{{Oesch} et~al.}{2010}]{Oesch2010}
{Oesch} P.~A., et~al.,  2010, \apjl, 714, L47


\bibitem[\protect\citeauthoryear{{Poggianti} et~al.,}{{Poggianti}
  et~al.}{2006}]{poggianti2006}
{Poggianti} B.~M.,  et~al., 2006, \apj, 642, 188

\bibitem[\protect\citeauthoryear{{Poggianti}, {Smail}, {Dressler}, {Couch},
  {Barger}, {Butcher}, {Ellis} \& {Oemler}}{{Poggianti}
  et~al.}{1999}]{Poggianti1999}
{Poggianti} B.~M.,  {Smail} I.,  {Dressler} A.,  {Couch} W.~J.,  {Barger}
  A.~J.,  {Butcher} H.,  {Ellis} R.~S.,    {Oemler} A.~J.,  1999, \apj, 518,
  576

\bibitem[\protect\citeauthoryear{{Poggianti} \& {van Gorkom}}{{Poggianti} \&
  {van Gorkom}}{2001}]{PoggiantiVG2001}
{Poggianti} B.~M.,  {van Gorkom} J.~H.,  2001, in {J.~E.~Hibbard, M.~Rupen, \&
  J.~H.~van Gorkom} ed., Gas and Galaxy Evolution Vol.~240 of Astronomical
  Society of the Pacific Conference Series, {Environmental Effects on Gas and
  Galaxy Evolution in Clusters}.
p.~599

\bibitem[\protect\citeauthoryear{{Porter}, {Raychaudhury}, {Pimbblet} \&
  {Drinkwater}}{{Porter} et~al.}{2008}]{Porter2008}
{Porter} S.~C.,  {Raychaudhury} S.,  {Pimbblet} K.~A.,    {Drinkwater} M.~J.,
  2008, \mnras, 388, 1152

\bibitem[\protect\citeauthoryear{{Rines} \& {Diaferio}}{{Rines} \&
  {Diaferio}}{2006}]{RinesDiaferio2006}
{Rines} K.,  {Diaferio} A.,  2006, \aj, 132, 1275

\bibitem[\protect\citeauthoryear{{Roediger}}{{Roediger}}{2009}]{Roediger2009}
{Roediger} E.,  2009, Astronomische Nachrichten, 330, 888

\bibitem[\protect\citeauthoryear{{Roychowdhury}, {Chengalur}, {Chiboucas},
  {Karachentsev}, {Tully} \& {Kaisin}}{{Roychowdhury}
  et~al.}{2012}]{Roychowdhury2012}
{Roychowdhury} S.,  {Chengalur} J.~N.,  {Chiboucas} K.,  {Karachentsev} I.~D.,
  {Tully} R.~B.,    {Kaisin} S.~S.,  2012, \mnras, 426, 665

\bibitem[\protect\citeauthoryear{{Scott}, {Bravo-Alfaro}, {Brinks}, {Caretta},
  {Cortese}, {Boselli}, {Hardcastle}, {Croston} \& {Plauchu}}{{Scott}
  et~al.}{2010}]{Scott2010}
{Scott} T.~C.,  {Bravo-Alfaro} H.,  {Brinks} E.,  {Caretta} C.~A.,  {Cortese}
  L.,  {Boselli} A.,  {Hardcastle} M.~J.,  {Croston} J.~H.,    {Plauchu} I.,
  2010, \mnras, 403, 1175

\bibitem[\protect\citeauthoryear{{Scott}, {Cortese}, {Brinks}, {Bravo-Alfaro},
  {Auld} \& {Minchin}}{{Scott} et~al.}{2012}]{Scott2012}
{Scott} T.~C.,  {Cortese} L.,  {Brinks} E.,  {Bravo-Alfaro} H.,  {Auld} R.,
  {Minchin} R.,  2012, \mnras, 419, L19

\bibitem[\protect\citeauthoryear{{Smith}, {Kneib}, {Smail}, {Mazzotta},
  {Ebeling} \& {Czoske}}{{Smith} et~al.}{2005}]{Smith2005}
{Smith} G.~P.,  {Kneib} J.-P.,  {Smail} I.,  {Mazzotta} P.,  {Ebeling} H.,
  {Czoske} O.,  2005, \mnras, 359, 417

\bibitem[\protect\citeauthoryear{{Tonnesen} \& {Bryan}}{{Tonnesen} \&
  {Bryan}}{2009}]{TonnesenBryan2009}
{Tonnesen} S.,  {Bryan} G.~L.,  2009, \apj, 694, 789

\bibitem[\protect\citeauthoryear{{Treu}, {Ellis}, {Kneib}, {Dressler}, {Smail},
  {Czoske}, {Oemler} \& {Natarajan}}{{Treu} et~al.}{2003}]{Treu2003}
{Treu} T.,  {Ellis} R.~S.,  {Kneib} J.,  {Dressler} A.,  {Smail} I.,  {Czoske}
  O.,  {Oemler} A.,    {Natarajan} P.,  2003, \apj, 591, 53

\bibitem[\protect\citeauthoryear{{van der Hulst}}{{van der
  Hulst}}{1979}]{vanDerHulst1979}
{van der Hulst} J.~M.,  1979, \aap, 75, 97

\bibitem[\protect\citeauthoryear{{van Dokkum}, {Franx}, {Kelson},
  {Illingworth}, {Fisher} \& {Fabricant}}{{van Dokkum}
  et~al.}{1998}]{vanDokkum1998}
{van Dokkum} P.~G.,  {Franx} M.,  {Kelson} D.~D.,  {Illingworth} G.~D.,
  {Fisher} D.,    {Fabricant} D.,  1998, \apj, 500, 714

\bibitem[\protect\citeauthoryear{{Verdes-Montenegro}, {Yun}, {Williams},
  {Huchtmeier}, {Del Olmo} \& {Perea}}{{Verdes-Montenegro}
  et~al.}{2001}]{VerdesMontenegro2001}
{Verdes-Montenegro} L.,  {Yun} M.~S.,  {Williams} B.~A.,  {Huchtmeier} W.~K.,
  {Del Olmo} A.,    {Perea} J.,  2001, \aap, 377, 812

\bibitem[\protect\citeauthoryear{{Verheijen}, {van Gorkom}, {Szomoru},
  {Dwarakanath}, {Poggianti} \& {Schiminovich}}{{Verheijen}
  et~al.}{2007}]{Verheijen2007}
{Verheijen} M.,  {van Gorkom} J.~H.,  {Szomoru} A.,  {Dwarakanath} K.~S.,
  {Poggianti} B.~M.,    {Schiminovich} D.,  2007, \apjl, 668, L9


\bibitem[\protect\citeauthoryear{{Voges} et~al.,}{{Voges} et~al.}{1999}]{Voges99}
{Voges} W., et~al.,   1999, \aap, 349, 389


\bibitem[\protect\citeauthoryear{{Vollmer}}{{Vollmer}}{2003}]{Vollmer2003}
{Vollmer} B.,  2003, \aap, 398, 525

\bibitem[\protect\citeauthoryear{{Vulcani}, {Poggianti}, {Finn}, {Rudnick},
  {Desai} \& {Bamford}}{{Vulcani} et~al.}{2010}]{Vulcani2010}
{Vulcani} B.,  {Poggianti} B.~M.,  {Finn} R.~A.,  {Rudnick} G.,  {Desai} V.,
  {Bamford} S.,  2010, \apjl, 710, L1

\bibitem[\protect\citeauthoryear{{Wilman}, {Oemler} Jr., {Mulchaey}, {McGee},
  {Balogh} \& {Bower}}{{Wilman} et~al.}{2009}]{Wilman2009}
{Wilman} D.~J.,  {Oemler} Jr. A.,  {Mulchaey} J.~S.,  {McGee} S.~L.,  {Balogh}
  M.~L.,    {Bower} R.~G.,  2009, \apj, 692, 298

\end{thebibliography}
%\bibliographystyle{mn2e}

%\bibitem[\protect\citeauthoryear{{Bolzonella},et~al.,}{{Bolzonella} et~al.}{2010}]{Bolzonella2010}
%{Bolzonella} M., et~al., 2010,
%  \aap, 524, A76+

%\bibitem[\protect\citeauthoryear{{Giovanelli}, et~al.,}{{Giovanelli} et~al.} {2005}] {Giovanelli2005}
%{Giovanelli} R.,  et~al.,  2005, \aj, 130, 2598

%\bibitem[\protect\citeauthoryear{{Meyer} et~al.,}{{Meyer} et~al.} {2004}]{Meyer2004}
%{Meyer} M.~J.,  et~al.,  2004, \mnras, 350, 1195

%\bibitem[\protect\citeauthoryear{{Oesch} et~al.,}{{Oesch} et~al.}{2010}]{Oesch2010}
%{Oesch} P.~A., et~al.,  2010, \apjl, 714, L47

%\bibitem[\protect\citeauthoryear{{Serra} et~al.,}{{Serra} et~al.}{2012}]{Serra2012}
%{Serra} P.,  et~al.,  2012, \mnras, 422, 1835

%\bibitem[\protect\citeauthoryear{{Voges} et~al.,}{{Voges} et~al.}{1999}]{Voges99}
%{Voges} W., et~al.,   1999, \aap, 349, 389

%\end{thebibliography}

\newpage
\appendix

\newpage
\section{Other cluster/groups outside the surveyed volume}

In Table~\ref{structures} we presented the main characteristics of the cluster/groups found in the BUDHIES volume. For completeness and future reference we show here a list of other structures identified by our spectroscopic campaign outside the redshift range of the HI survey ($0.164\leqslant z \leqslant0.224$). 

\begin{table*}
\begin{center}
 \caption{Clusters, groups and other structures identified in the fields of A963 and A2192 outside the redshift range of the HI-observations.} 
\label{structures_2}
\begin{tabular}{llccccl}
\hline\\[-1mm]
Field 	& Name of structure   &  $z_{c}$   & No. members	& $\sigma_{\rm cl}$ & R$_{200}$ & Remarks\\ 
 	& 		      &                      & ($R<19.4$mag)	& (km s$^{-1}$) 	    & ($h^{-1}$ Mpc)	 & \\ 
\hline\\[-2mm]
A963	& \textbf{A963\_4}  & 0.1478		& 24	& 290$\pm53$	&0.47	& Sparse in $\alpha$ and $\delta$	\\
\hline\\[-2mm]
A2192	& \textbf{A2192\_4} & 0.1581	& 26	& 454$\pm64$	&0.73	 & Well-defined group.	\\
	& \textbf{A2192\_5} & 0.1348		& 35	& 717$\pm91$	&1.16	& Scattered, double peaked in z	\\
	& \textbf{A2192\_6} & 0.2319		& 26	& 579$\pm92$	&0.89	&``companion'' of A2192\_3	\\
\hline\\
\end{tabular}  
\\ 
\end{center}
\end{table*}

\newpage

\section{The Dressler-Shectman test: Bubble Plots}
\label{ap:DS}

\begin{figure*}
\begin{center}
  \includegraphics[width=0.49\textwidth]{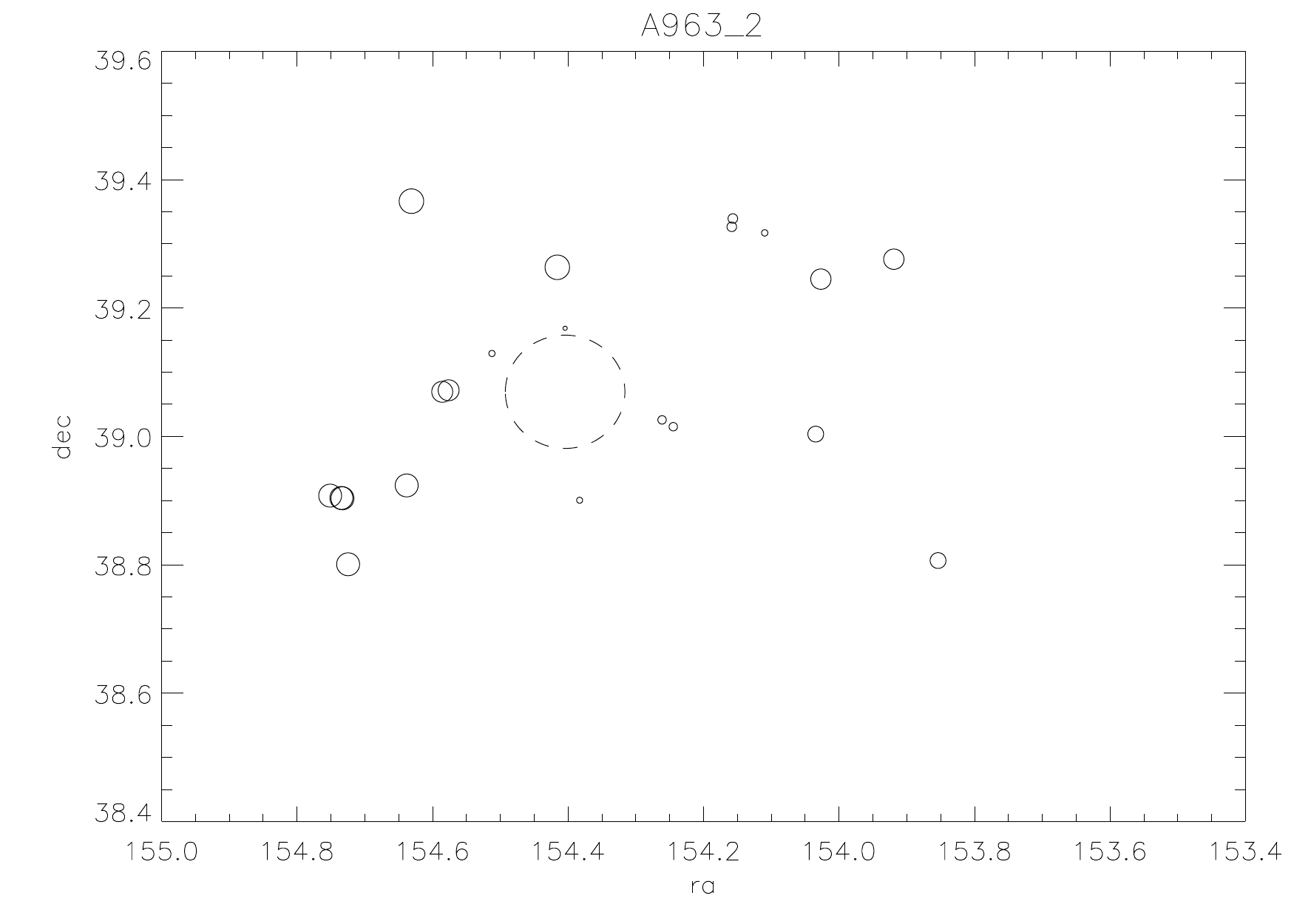}
  \includegraphics[width=0.49\textwidth]{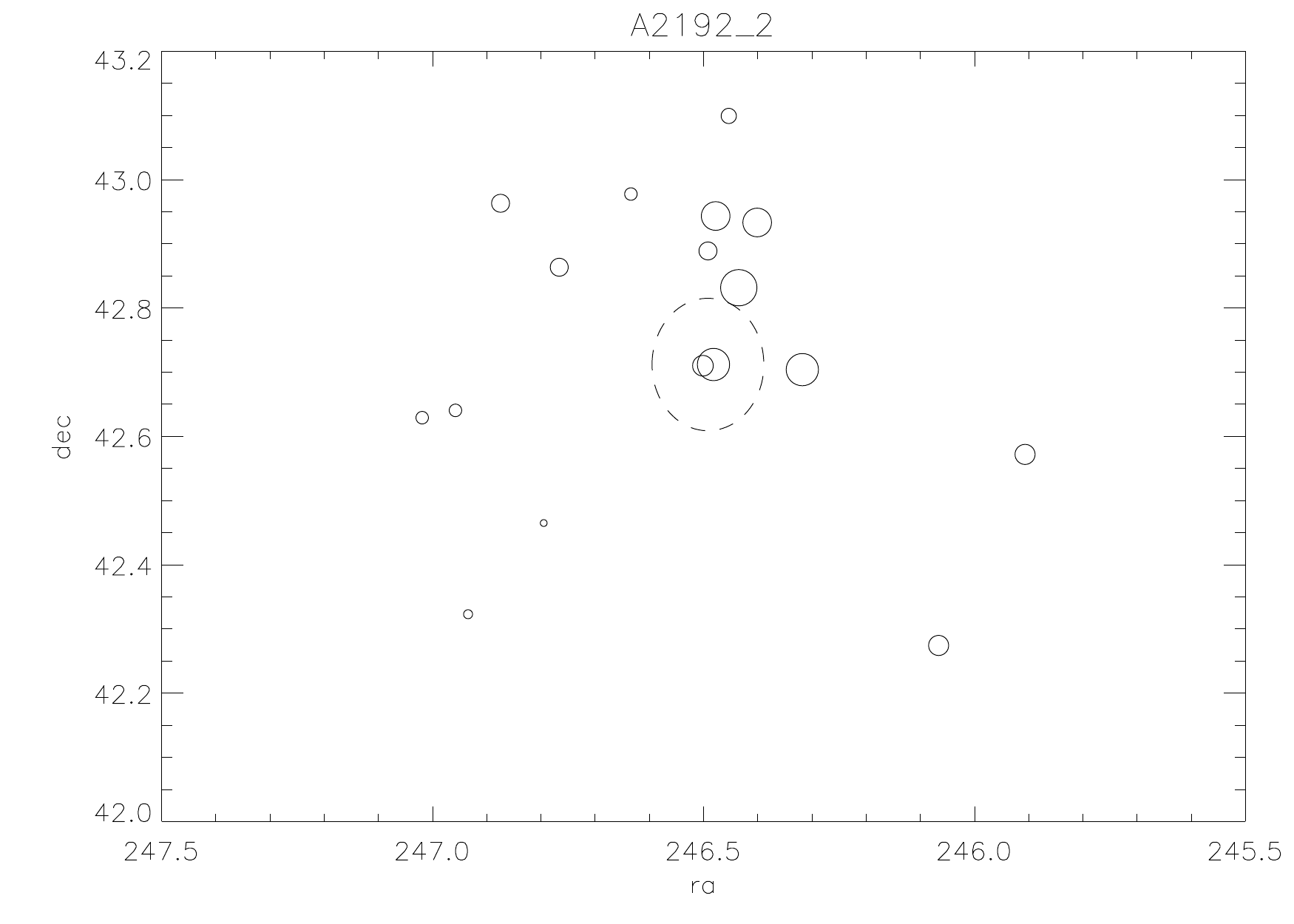}
  \includegraphics[width=0.49\textwidth]{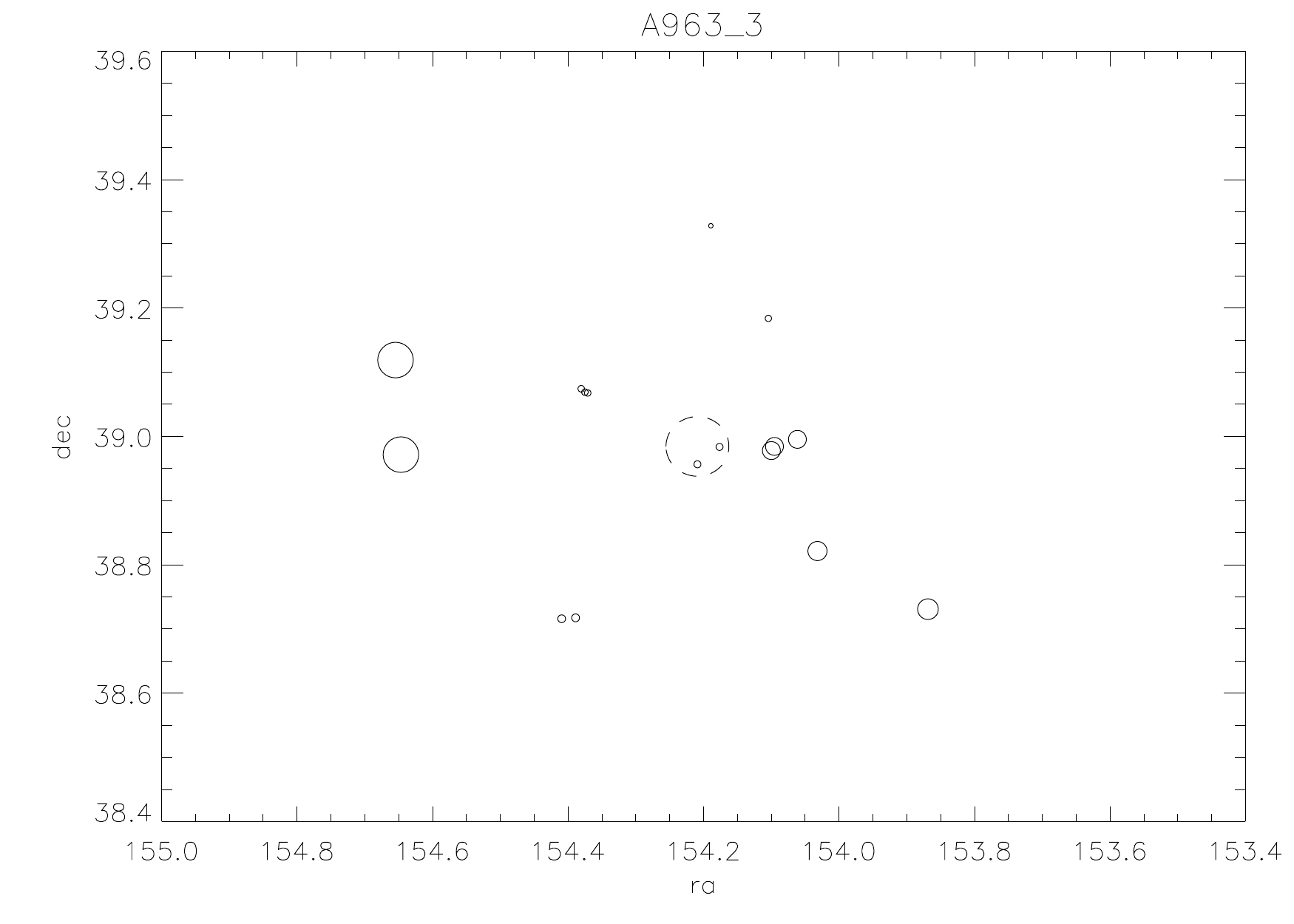}
  \includegraphics[width=0.49\textwidth]{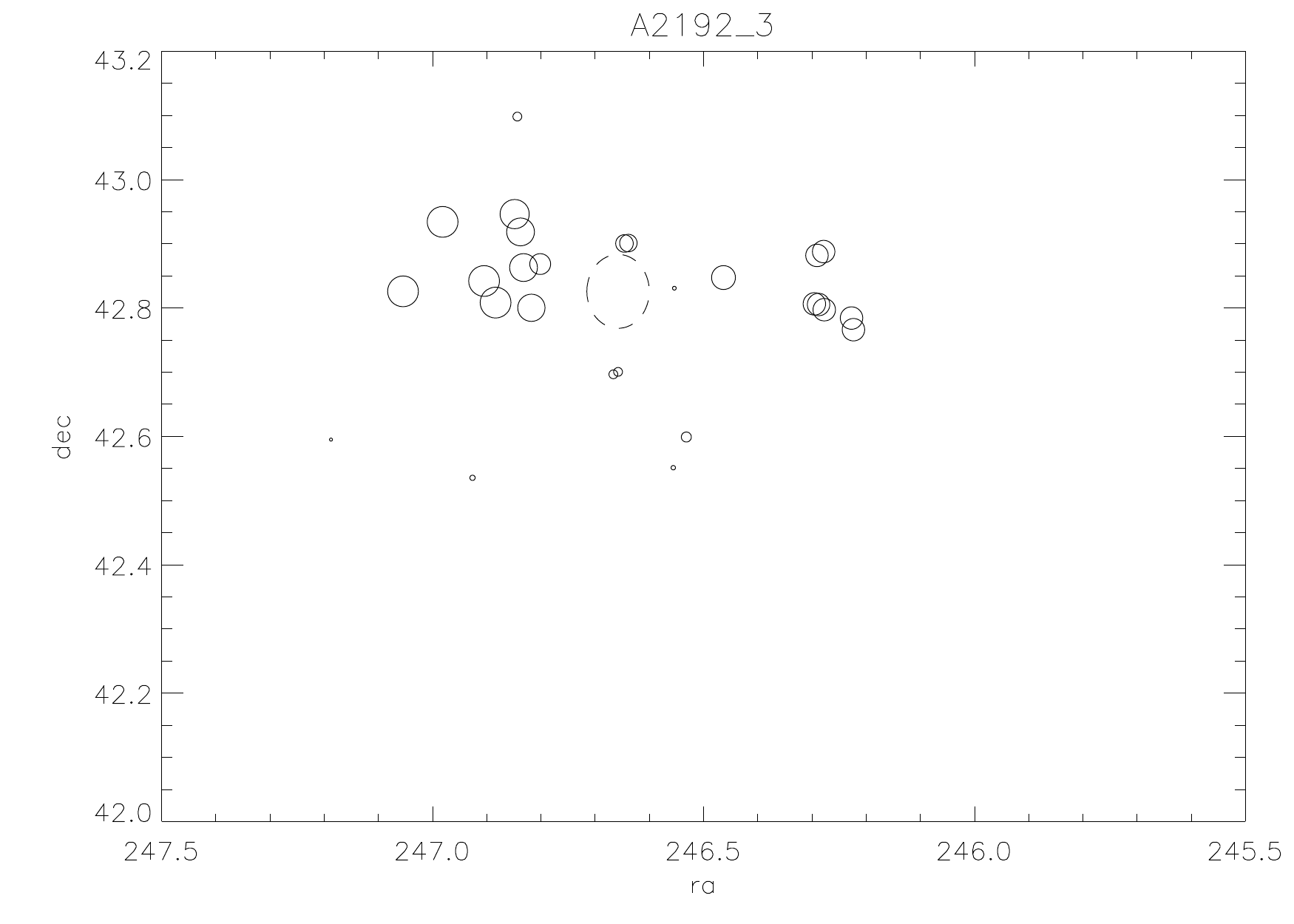}
\end{center} 
\caption{Figure~\ref{DS_bubble_plots} showed the  \citet{DresslerShectman1988} ``bubble-plot'' for the two main clusters in our sample. We show here the remaining cluster/groups. Galaxies are plotted with circles with a diameter that scales with  exp($\delta_{i}$).  Dashed circles indicate the $R_{200}$ radius of the clusters. Only galaxies with good quality redshifts were considered.}
 \label{DS_bubble_plots_cont}
\end{figure*}

\newpage

\section{Spectroscopic catalogue}
\label{ap:table}

In the following, we show an example 10 rows of the spectroscopic catalogue associated with this paper. 

\begin{table*}
\begin{center}
 \caption{This table contains the optical information for the galaxies observed with the WHT for which a redshift was obtained. Only 10 example rows are shown here. The full table is available in the online version of the paper. The ID is preceded by ``IJ'' and contains the right ascention and declination of the INT photometry in hms and dms format. The redshift quality ranges from 1 to -2, only redshifts quality $\geq -1$ are to be trusted. The type of spectra is also indicated: ``em'' stands for emission-line spectra, ``ab'' for absorption, ``em+ab'' means that there are both types of features clearly visible, and ``nan'' corresponds to galaxies with unclear spectral features. } 
\label{spec_table}
\begin{tabular}{lccccccc}
\hline\\[-1mm]
Field 	& ID	& $R$	& $B$ 	& redshift 	& redshift  	& type of	&	 EW[OII] 	\\ 
 	& 	& mag	& mag 	&  		& quality 	& spectra	&	  	\\ 
  \hline\\
A963 	& IJ101527.90+390603.6		&    18.67	&  20.03	&  0.2054	& 1	& em+ab	& 14.28	\\ 
A963 	& IJ101544.32+384910.9		&    19.99	&  20.78	&  0.2064	& 1	& em/ab	&  11.1	\\ 
A963 	& IJ101532.79+384805.1		&    18.45	&  20.71	&  0.2060	& 1	&  ab	&  $--$	\\ 
A963 	& IJ101659.58+384956.2		&    18.78	&  20.82	&  0.2066	& -1	& ab	&  $--$	\\ 
A963 	& IJ101721.29+390602.5		&    18.47	&  20.35	&  0.1993	& 1	& nan 	&  $--$	\\ 
A963 	& IJ101701.54+390009.3		&    19.77	&  20.85	&  0.1650	& 1	&  em	& 57.22	\\ 
A963 	& IJ101534.81+390912.0		&    18.20	&  20.42	&  0.1749	& 1	& ab 	& $--$ 	\\ 
A2192 	& IJ162431.50+424621.7		&    18.29	&  19.93	&  0.2612	& 1	 & ab 	& $--$ 			\\ 
A2192 	& IJ162436.20+424111.7		&    19.97	&  20.91	&  0.1427	& -2	 &  ab	&  $--$	\\ 
A2192 	& IJ162658.84+421557.1		&    19.34	&  21.20	&  0.1783	& -1	 &  ab	& $--$ 	\\ 
\hline\\[1mm]
\end{tabular}  
\\ 
\end{center}
\end{table*}

\end{document}